# Transformation and amplification of light modulated by a traveling wave with a relatively low frequency

M. Sumetsky

Aston Institute of Photonic Technologies, Aston University, Birmingham B4 7ET, UK
*e-mail*: m.sumetsky@aston.ac.uk

The behavior of electromagnetic waves in a medium modulated in time and space, largely investigated decades ago, has recently attracted renewed interest. Here, we solve an intriguing problem of this research: can light with an initial frequency $\omega_0$ be amplified in a realistic photonic circuit solely pumped by a traveling wave with a much lower frequency $\omega_p \ll \omega_0$? Generally, the bandwidth of the modulation-induced optical frequency comb spectrum can be substantially broadened when the phase velocity of the traveling wave, $v_p$, approaches the phase velocity of light, $v_0$. However, in realistic photonic waveguides, the amplification effect remains small due to the unfeasible modulation and waveguide parameters required. In contrast, we demonstrate that modulating an optical resonator with a traveling wave that has a small phase velocity $v_p \ll v_0$ (rather than a synchronous $v_p \cong v_0$) can result in narrow-band light amplification, which is dramatically enhanced near the Brillouin phase-matching condition $\omega_p/v_p \cong 2\omega_0/v_0$. Our calculations show that the proposed amplifier of light can be realized in a lithium niobate racetrack resonator with millimeter-scale perimeter modulated by a surface acoustic wave with surprisingly small and practically achievable amplitude.

## I. INTRODUCTION

The propagation of electromagnetic waves along transmission lines, waveguides, and more complex electromagnetic and photonic circuits and structures *parametrically modulated in time and space* has been studied both theoretically and experimentally several decades ago [1-14] and has attracted recent interest [15-22]. The long-lasting interest in this area of research is caused by fascinating transformations and features exhibited by a wave in a time-modulated medium – not possible in the stationary regime – and their ongoing and potential applications. These features include frequency conversion, nonreciprocal behavior, dynamic bandgaps, and amplification.

Frequency conversion of waves inherently occurs in all cases of their time-modulation [1-22] unless the modulation is too weak or adiabatic. One of the essential practical outcomes of investigation of time-modulated photonics circuits was the ongoing development of optical modulators [23-28]. Numerous earlier and recent papers investigated modulation-induced frequency comb (OFC) generation [29, 30] and sideband transitions including the effects of complete inelastic transparency and propagation nonreciprocity [31-41]. Temporal modulation can also create dynamic bandgaps, preventing certain frequencies from propagating and causing dynamic localization, where waves are trapped in localized regions due to constructive interference [19, 42, 43]. Additionally, modulating the properties of a medium can affect the group velocity of waves, resulting in slow or fast light [44]. Temporal modulation also allows for real-time wavefront control, enabling dynamic beam steering, focusing, and diffraction pattern manipulation, which are important in adaptive optics and beamforming technologies [45].

A crucial feature of wave propagation in a time-modulated medium is the potential for amplification. Modulation can transfer energy to the wave, enhancing its amplitude, or extracting energy from it, leading to attenuation. For example, temporal modulation of the medium refractive index $\Delta n(t) = \Delta n_p \cos(\omega_p t)$ with frequency $\omega_p$ close to a multiple of the input electromagnetic wave half-frequency, $\omega_0/2$, can lead to amplification or attenuation of this wave described by Floquet theory ( see e.g., [9, 46]). For applications in optics, the amplification is customary achieved by pumping with a high-power light which frequency $\omega_p$ is *comparable* to the frequency of input light to be amplified, $\omega_p/\omega_0 \sim 1$ [47]. For example, in Brillouin and Raman lasers, the acoustic and molecular vibrations are excited by a pump light with a frequency $\omega_p$ that is close to or comparable with the frequency of amplified light $\omega_0$.

However, is it possible to amplify an optical wave with frequency $\omega_0$ in a realistic photonic circuit modulated *solely by a travelling wave with much smaller frequency $\omega_p \ll \omega_0$* (e.g., by an acoustic or RF wave) in the absence of pumping light? For the ideal case of a waveguide with *negligible dispersion and losses over the large bandwidth $\Delta \omega_B \gtrsim \omega_0$*, a positive answer to this question was given several decades ago [5, 6, 13]. The authors of Ref. [5, 6] (and, independently, the authors of Ref. [48]) found the exact solution of this problem for a one-dimensional propagation of a wave in a medium with constant impedance modulated by a traveling wave [5] and its asymptotic (eikonal, WKB) solution for a medium with constant permeability [6, 48]. It was shown that the amplification under these conditions is indeed possible if the phase velocity of light $v_0$ is close to the phase velocity $v_p$ of the traveling wave. These results are irrelevant to realistic photonic circuits since their transmission loss and dispersion are *never negligible within the frequency bandwidth $\Delta \omega_B \gtrsim \omega_0$* required for the observation of substantial amplification. Consequently, the intriguing question if the light amplification can be achieved by modulating a photonic circuit with a low frequency $\omega_p \ll \omega_0$ remains open.

Here, we propose an answer to this question. First, we explore the described problem using the eikonal approximation, which is valid when modulation is slow in time and space, i.e., when it has a relatively small frequency $\omega_p \ll \omega_0$ (assumed throughout the paper), and wavenumber, $k_p = \frac{\omega_p}{v_p} \ll k_0 = \frac{\omega_0}{v_0}$ [13, 14]. We show that, due to the requirement of the broadband and lossless transmission [5, 6], the substantial amplification of light by a low frequency modulation is unfeasible in realistic optical waveguides if the phase velocities of modulating wave and light have the same order, $v_p \sim v_0$, and, in particular, are equal to each other, $v_p = v_0$. For a very small phase velocity $v_p$, when the wave numbers of light and modulating wave become comparable, $\frac{\omega_p}{v_p} \sim \frac{\omega_0}{v_0}$, the eikonal approximation fails, but a regular perturbation theory over the modulation amplitude of refractive index comes into force. Using this theory, we show that the amplification effect can significantly increase with a decrease of the phase velocity of modulating wave, $v_p$. However, it still remains small for realistic waveguides and feasible modulation methods. In contrast, we demonstrate that light propagating through an optical racetrack lithium niobate resonator with realistic characteristics can be significantly amplified by an acoustic wave with surprisingly small and practically achievable amplitude. The amplification occurs within a narrow bandwidth $\Delta\omega_B$ near the Brillouin phase matching condition $\omega_p/v_p \cong 2\omega_0/v_0$.

## II. TRANSFORMATION AND AMPLIFICATION OF LIGHT IN AN OPTICAL WAVEGUIDE IN THE EIKONAL APPROXIMATION

In this section, we consider the propagation of an optical wave along a dispersionless waveguide modulated by a low frequency traveling wave illustrated in Fig. 1 in the eikonal approximation. We assume that the input wave has the initial phase velocity $v_0$ and frequency $\omega_0$, while the modulating traveling wave has the phase velocity $v_p$ and much smaller frequency $\omega_p \ll \omega_0$. The one-dimensional wave propagation is described by the wave equation

$$\left(n^2 E\right)_{tt} - c^2 E_{xx} = 0, \qquad (1)$$

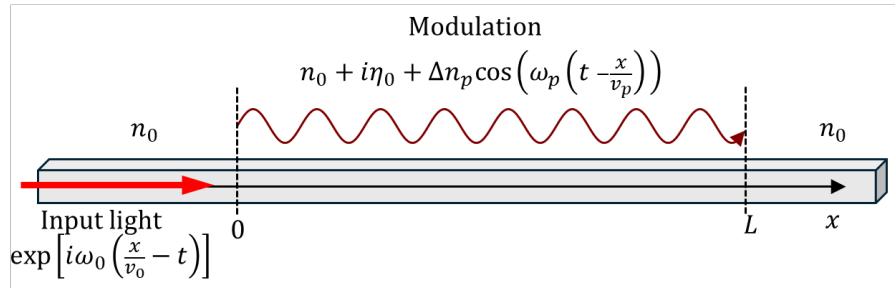

**Fig. 1.** An optical waveguide with the refractive index modulated by a traveling wave along the interval $(0, L)$.

where the subindices denote partial derivatives, and $c$ is the speed of light. The dependence of refractive index on time $t$ and coordinate $x$ along the waveguide is set to

$$n(x,t) = n_0 + \Delta n(x,t),$$

$$\Delta n(x,t) = \begin{cases} i\eta_0 + \Delta n_p \cos\left(\omega_p \left(t - \frac{x}{v_p}\right)\right) & 0 < x < x_p \\ 0 & \text{elsewhere} \end{cases}. \qquad (2)$$

In this equation, $\eta_0$ defines the propagation loss. The boundary condition for the solution of Eq. (1) along a waveguide (Fig. 1) is defined by the input wave with frequency $\omega_0$ and phase velocity $v_0 = c/n_0$:

$$E^{(in)}(x,t) = \exp\left[i\omega_0\left(\frac{x}{v_0} - t\right)\right], \quad v_0 = \frac{c}{n_0}, \quad x < 0. \qquad (3)$$

We notice that even for a small modulation amplitude $\Delta n_p \ll n_0$, solution of Eq. (1) by the perturbation theory over $\Delta n_p$ is incorrect if the phase velocities $v_p$ and $v_0$ are close to each other so that $|v_p - v_0|/v_0 \sim \Delta n_p/n_0 \ll 1$ [49]. Alternatively, the propagation of waves in a media whose parameters are slowly varying in space and time can be described in the eikonal approximation [5, 6, 13, 14], also known as the WKB and semiclassical approximation in quantum mechanics [50, 51] and the geometric optics approximation in the electromagnetic theory [14].

### A. Solution of the wave equation in the eikonal (WKB) approximation

Application of the eikonal approximation requires that the parameters of the optical waveguide vary slowly both in time and space. Specifically, the frequency $\omega_p$ and wavenumber $k_p = \omega_p/v_p$ of the traveling wave should be small compared to the frequency $\omega_0$ and wavenumber $k_0 = \omega_0/v_0$ of the input wave:

$$\frac{\omega_p}{\omega_0} \ll 1, \quad \frac{k_p}{k_0} = \frac{\omega_p}{v_p}\frac{v_0}{\omega_0} \ll 1. \tag{4}$$

We also assume that the material waveguide loss $\eta_0$ is relatively small, $\eta_0 \ll n_0$. Then, the solution of Eq. (1) $E(x,t)$ in the region $0 < x < L$ can be found by the eikonal (semiclassical) theory [13, 14]. In this theory, the solution of Eq. (1) is presented as $E(x,t) = \exp(iS_0(x,t)/\varepsilon)\sum_{n=0}^{\infty} \varepsilon^n U_n(x,t)$ where the small parameter $\varepsilon = \max(\omega_p/\omega_0, k_p/k_0) \ll 1$ and $S(x,t) = S_0(x,t)/\varepsilon$ is the eikonal satisfying equation $n^2(x,t)S_t^2 - c^2 S_x^2 = 0$. For the small modulation, $\Delta n_0 \ll n_0$, or, alternatively, for the modulation adiabatically switching on near the coordinate $x = 0$ and off near $x = L$ (the switching region is not illustrated in Fig. 1), we can ignore the reflected waves at $x = 0$ and $x = L$. Then, similar to calculations of Ref. [6], we find the zero-order in $\varepsilon$ asymptotic solution of Eq. (1) with the refractive index defined by Eq. (2) and the boundary condition of Eq. (3) (see Appendix A):

$$E(x,t) = U_0(x,t)\exp(iS(x,t)),$$

$$S(x,t) = -\omega_0 \bar{t}(\xi(x,t)), \tag{5}$$

$$U_0(x,t) = \frac{\left(1+\mu\cos\left(\omega_p \bar{t}(\xi(x,t))\right)\right)\sqrt{1+\mu\delta\cos\left(\omega_p \bar{t}(\xi(x,t))\right)}}{\left(1+\mu\cos\left(\omega_p\left(t-\frac{x}{v_p}\right)\right)\right)\sqrt{1+\mu\delta\cos\left(\omega_p\left(t-\frac{x}{v_p}\right)\right)}},$$

$$\bar{t}(\xi) = -\frac{2}{\omega_p}\Xi\left(\sqrt{\frac{1+\mu}{1-\mu}}, \frac{\sqrt{1-\mu^2}}{2v_0 v_p}(v_p - v_0)\omega_p \xi\right), \tag{6}$$

$$\xi(x,t) = x - \frac{2v_0 v_p}{\omega_p(v_p-v_0)\sqrt{1-\mu^2}}\Xi\left(\sqrt{\frac{1-\mu}{1+\mu}}, \frac{\omega_p}{2}\left(t-\frac{x}{v_p}\right)\right), \tag{7}$$

$$\mu = \frac{\Delta n_p}{n_0 \delta}, \quad \delta = \left(1 - \frac{v_0}{v_p} + i\frac{\eta_0}{n_0}\right). \tag{8}$$

Here, function $\Xi(x,y)$ is the smooth continuation of $\arctan(x \cdot \tan(y))$ as a function of $y$ (see Eq. (A12) in Appendix A).

It follows from Eqs. (5)-(8) that the field $E(x,t)$ is periodic in time with the period $2\pi/\omega_p$ though may be aperiodic in space. In major calculations below we will ignore the factors under the square roots in Eq. (5) since we always assume that $\mu\delta = \Delta n_p/n_0 \ll 1$. The behavior of solution $E(x,t)$ at position $x$ is characterized by the complex valued (in the presence of losses) *synchronization parameter* $\mu$ introduced in Eq. (8). For small losses considered below, $\eta_0 \ll \Delta n_p$, this parameter is the ratio of the relative modulation amplitude $\Delta n_p/n_0$ and the relative proximity of velocities of the input wave and the modulating traveling wave, $1 - v/v_p$. Following Ref. [5, 6], we call the modulation *asynchronous* if $|\mu| < 1$, call it *completely asynchronous* if $|\mu| \ll 1$, (in particular, call it *instantaneous* if $v_p = \infty$), call it *synchronous* if $|\mu| > 1$, and call it *completely synchronous* if $|\mu| \gg 1$ (in particular, if $v_p = v$). While the solution defined by Eqs. (5)-(8) is quasiperiodic in space in the case of asynchronous modulation, it can exponentially grow in space for synchronous modulation.

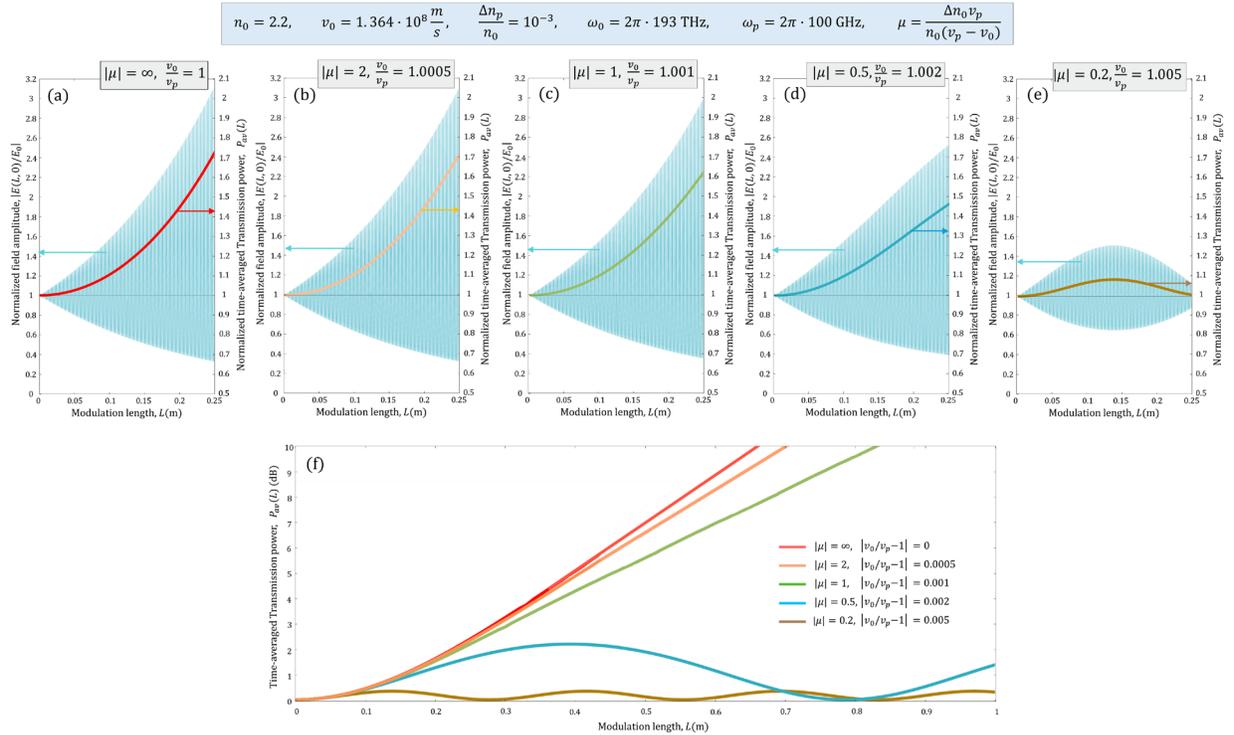

**Fig. 2.** The amplitude and time-averaged power of the output wave as a function of modulation length $L$ for different synchronization parameters $\mu$ corresponding to close phase velocities $v$ and $v_p$ of the input light and modulation. (a)-(e) The amplitude (left vertical axes) and normalized time-averaged power (right vertical axes) for modulation lengths $0 < L < 0.25$ m. (a) $|\mu| = \infty, v = v_p$ (completely synchronous case); (b) $|\mu| = 2, v = 1.001v_p$ (synchronous case); (c) $|\mu| = 1, v = 1.001v_p$; (d) $|\mu| = 0.5, v = 1.002v_p$ (asynchronous case); (e) $|\mu| = 0.2, v = 1.005v_p$ (asynchronous case); (f) Time-averaged power $P_{av}(L)$ for modulation lengths $0 < L < 1$ m for the relations between $v$ and $v_p$ of plots (a)-(e).

## B. The wave amplification effect in the absence of losses

Under the condition of negligible material losses, $\eta_0 = 0$, the synchronization parameter $\mu$ is real and there exist two qualitatively different cases of the spatially unstable ($|\mu| > 1$) and spatially stable ($|\mu| < 1$) solutions, corresponding to the synchronous and the asynchronous cases introduced above. Fig. 2 presents the characteristic behavior of the normalized field amplitude $|U_0(L,t)|$ at a fixed time $t = 0$ and time-averaged normalized wave power, an average of the squared amplitude $U_0(x,t)$ defined by Eq. (5) over the time period $2\pi/\omega_p$:

$$P_{av}(L) = \frac{\omega_p}{2\pi} \int_0^{2\pi/\omega_p} U_0(L,t)^2 \, dt. \qquad (9)$$

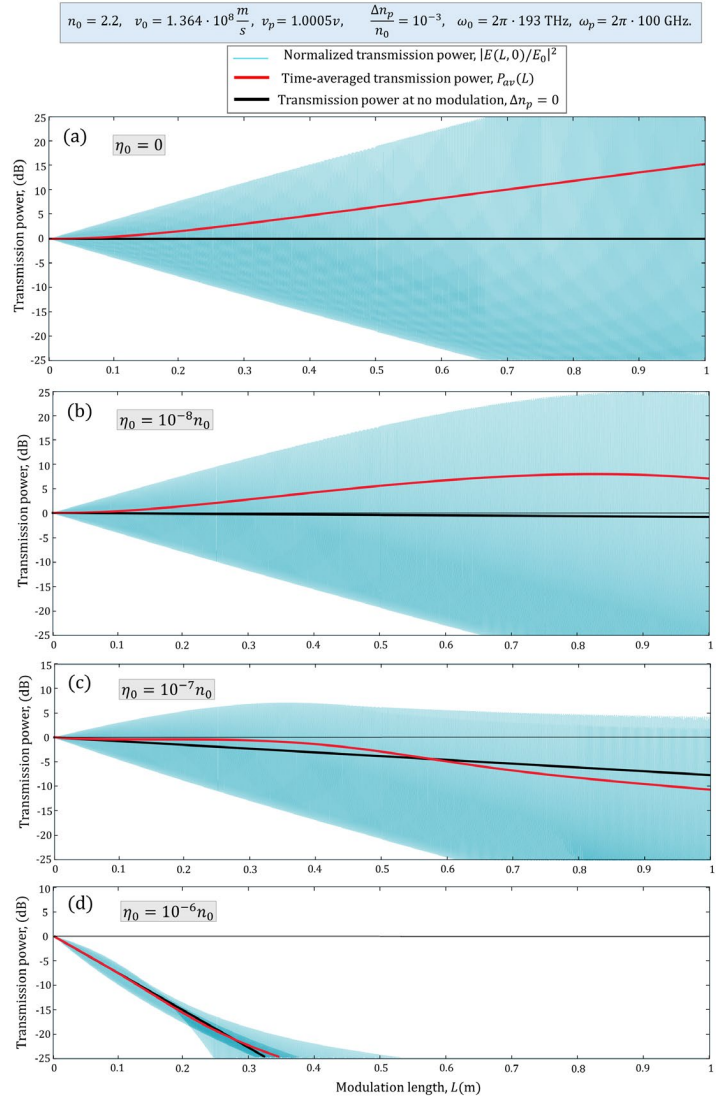

**Fig. 3.** The transmission power and time-averaged power of the output wave as a function of modulation length $L$ for different losses $\eta_0$ in the synchronous case $|\mu| = 0.2$, $v_p = 1.0005v$. (a) $\eta_0 = 0$. (b) $\eta_0 = 10^{-8}n_0$. (c) $\eta_0 = 10^{-7}n_0$. (d) $\eta_0 = 10^{-6}n_0$.

Fig. 2(f) presents $P_{av}(L)$ as a function of modulation length $L$ for different synchronization parameters $\mu$. In this figure, we consider the propagation of light along the lithium niobate waveguide with refractive index $n_0 = 2.2$ modulated with relative amplitude $\Delta n_p/n_0 = 10^{-3}$. The input light frequency and modulation frequency are set to $\omega_0 = 2\pi \cdot 193$ THz and $\omega_p = 2\pi \cdot 100$ GHz [52]. The purpose of so large modulation amplitude $\Delta n_p$ and frequency $\omega_p$ considered is to evaluate the *largest possible effects of modulation* including the largest possible amplification. To visually resolve the fine spatial oscillations of the wave amplitude, Figs. 2(a)-(e) show the behavior of $|U_0(L, 0)|$ (blue frequently oscillating curves) and

$P_{av}(L)$ (bold curves of different color) along the interval $0 < L < 0.25$ m, while Fig. 2(f) shows the behavior of $P_{av}(L)$ over a longer interval $0 < L < 1$ m. It is seen that for $|\mu| > 1$ the wave amplitude $|U_0(L,t)|$ oscillates and grows with $L$ while its time-averaged value grows exponentially. In contrast, for $|\mu| < 1$, both $|U_0(L,0)|$ and $P_{av}(L)$ remain, respectively, quasiperiodic and periodic as a function of modulation length $L$.

It is also seen from Fig. 2(f) that, for the parameters considered, the dependencies of the wave amplitude $|U_0(L,0)|$ and the time-averaged power $P_{av}(L)$ on $L$ are similar for modulation lengths $L < 0.1$ m. Consequently, in this interval, the proximity to the full synchronization condition $v_p = v$ does not enhance the wave amplification, which is always small at these modulation lengths.

## C. The effect of losses

Material losses can significantly modify the behavior of the propagating wave shown in Fig. 2. Here, we find the effect of relatively small though practically feasible material losses for an ideally dispersionless waveguide. We note that the light propagation loss

$$\alpha = \frac{\eta_0 \omega_0}{c} \tag{10}$$

of a lithium niobate waveguide can be as small as 0.2 dB/m [53, 54] corresponding to $\eta_0 \sim 10^{-8}$ and commonly has the order of 10 dB/m or greater [26, 27]. In Fig. 3 we consider the effect of broadband dispersionless material losses for the phase velocity relation $v_p = 1.0005 v_0$. In this figure, the blue curves show the normalized wave power $|U_0(L,0)|^2$ as a function of modulation length at a fixed time, $t = 0$, the black curves show this dependence for the unmodulated waveguide, $\Delta n_p = 0$, and the red curves are the dependencies of time-averaged wave power $P_{av}(L)$ on the modulation length $L$. We notice that the relation $v_p = 1.0005 v_0$ considered in Fig. 3 corresponds to the synchronization parameter $|\mu| = 2$ similar to that for the relation $v_0 = 1.0005 v_p$ considered in Fig. 2(b). Consequently, the behavior of the field power shown in Fig. 2(b) and the corresponding time-averaged field power (orange curve in Fig. 2(f)) is similar to those in Fig. 3(a) for $\eta_0 = 0$. We find that the effect of attenuation for $\eta_0 = 10^{-8}$ (Fig. 3(b)) is small for modulation lengths $L < 0.3$ m and grows for larger $L$. This effect is much stronger for $\eta_0 = 10^{-7}$ (Fig. 3(b)) and for $\eta_0 = 10^{-6}$ (Fig. 3(c)).

## D. Wave propagation and wave spectrum in the lossless completely synchronous case $v_p = v$

Of special interest is the ideal lossless and completely synchronous case when the phase velocities $v$ and $v_p$ are equal, $v_p = v$, illustrated in Fig. 2(a) and (f). For optical wave propagation, this case is also referred to as the luminal case [17]. Then, the expressions for the solution phase (eikonal) $S(x,t)$ and normalized amplitude $U_0(x,t)$ are simplified (see Appendix B):

$$E(x,t) = U_0(x,t) \exp(iS(x,t)),$$

$$S(x,t) = \frac{2\omega}{\omega_p} \arctan \left( \frac{\tanh\left(\frac{\Delta n_p \omega_p x}{2 n_0 v_0}\right) - \tan\left(\frac{\omega_p}{2}\left(t - \frac{x}{v_0}\right)\right)}{1 - \tanh\left(\frac{\Delta n_p \omega_p x}{2 n_0 v_0}\right) \tan\left(\frac{\omega_p}{2}\left(t - \frac{x}{v_0}\right)\right)} \right), \tag{11}$$

$$U_0(x,t) = \frac{1}{\cosh\left(\frac{\Delta n_p \omega_p x}{n_0 v_0}\right) - \sin\left(\omega_p\left(t - \frac{x}{v_0}\right)\right) \sinh\left(\frac{\Delta n \omega_p x}{n_0 v_0}\right)}.$$

Here, the expression for $U_0(x,t)$ is derived under the commonly satisfied condition $\Delta n_0/n_0 \ll 1$. For a relatively small argument of cosh (...) in the expression for $U_0(x,t)$ in Eq. (11), $x\Delta n_0\omega_p/n_0 v \ll 1$ (weak amplification), the expansion of $U_0(x,t)$ up to the second order in $x\Delta n_0\omega_p/n_0 v$ coincides with that found in Ref. [17] where the case $|\mu| \ll 1$ was considered (see below). From Eqs. (9) and (11), the time-averaged wave power can be found analytically:

$$P_{av}(L) = \cosh\left(\frac{\Delta n_p \omega_p L}{c}\right). \tag{12}$$

This result coincides with that found for the exactly solvable problem of waveguides with constant impedance [5]. It is seen from Fig. 2 that the completely synchronous condition corresponds to the maximum wave power amplification. It follows from Eq. (12) and also seen in Fig. 2(f) that the average amplification power grows exponentially with modulation length $L$ if $L \gg c/\Delta n_p \omega_p$. For the optical waveguide and modulation parameters indicated in Fig. 3 we have $L \cong 0.5$ m. We find from Eq. (11) that for $x \gg c/\Delta n_p \omega_p$ the amplitude $U_0(x,t)$ becomes a fast function of the coordinate and time if $\sin\left(\omega_p\left(t-\frac{x}{v_p}\right)\right)$ is close to unity. In the latter case the eikonal approximation may fail since its condition of validity reads (see Appendix B):

$$\frac{\omega_p}{\omega_0} \ll \exp\left(-\frac{2\Delta n_p \omega_p}{n_0 v_0}x\right). \tag{13}$$

This condition is well satisfied for the parameters considered in our numerical modelling.

For a relatively small amplification length $L \ll c/\Delta n_0 \omega_p$, the spectrum of the output light is localized near the input frequency $\omega_0$. Consequently, it is convenient to introduce the spectrum centered at $\omega_0$ by the expansion:

$$E(L,t) = \sum_{m=-\infty}^{\infty} U_m^{(c)} \exp(-i\omega_0 t + im\omega_p t), \tag{14}$$

$$U_m^{(c)} = \frac{\omega_p}{2\pi} \int_0^{2\pi/\omega_p} E(L,t) \exp(i\omega_0 t - im\omega_p t) dt. \tag{15}$$

For eikonal $S(x,t)$ and amplitude $U_0(x,t)$, which are slow functions of time, the integral for $U_n^{(c)}$ in Eq. (15) can be calculated by the stationary phase method. Calculations detailed in Appendix C show that the stationary points of this integral exist only within the frequency band

$$\omega_{B1}(L) < \omega < \omega_{B2}(L), \quad \omega_{B1,2} = \omega_0\left(2 - \exp\left(\pm\frac{\Delta n_p \omega_p}{n_0 v_0}L\right)\right) \tag{16}$$

with the bandwidth

$$\Delta \omega_B = \omega_{B2}(L) - \omega_{B2}(L) = 2\omega_0 \sinh\left(\frac{\Delta n_p \omega_p}{n_0 v_0}L\right). \tag{17}$$

Close to the edges of this band, the second derivative $S_{xx}$ tends to zero and the stationary phase method fails. The absence of the real stationary points of the integral in Eq. (15) outside of this frequency band suggests that the position of the spectral bandwidth of the solution given by Eq. (11) is determined by Eqs. (16) and (17). This result is confirmed by numerical calculations of the spectrum for different modulation lengths $L = 0.01, 0.05, 0.1, 0.2,$ and $0.5$ m presented in Fig. 4. For a relatively small modulation length $L = 0.01$ m, the average field amplification is negligible, $P_{av}(L) = 1.001$, and the spectral bandwidth is small compared to the input light frequency, $\Delta v_B = \Delta \omega_B/2\pi = 18$ THz $\ll \omega_0/2\pi = 193$ THz. At $L = 0.05$ m and $0.1$ m, the bandwidth $\Delta v_B$ becomes comparable to the input frequency, though the average amplification remains small. For larger modulation lengths $x_p$, the left-hand-side bandwidth edge $\omega_{B1}(L)$ becomes negative and exponentially grows with $L$. Alternatively, for large $L$ the right-hand-side edge $\omega_{B2}(L)$ tends to $2\omega_0$ and the output wave spectrum localizes in the vicinity of $2\omega_0$ as illustrated by the spectrum of the output wave at $L = 0.5$ m in Fig. 4.

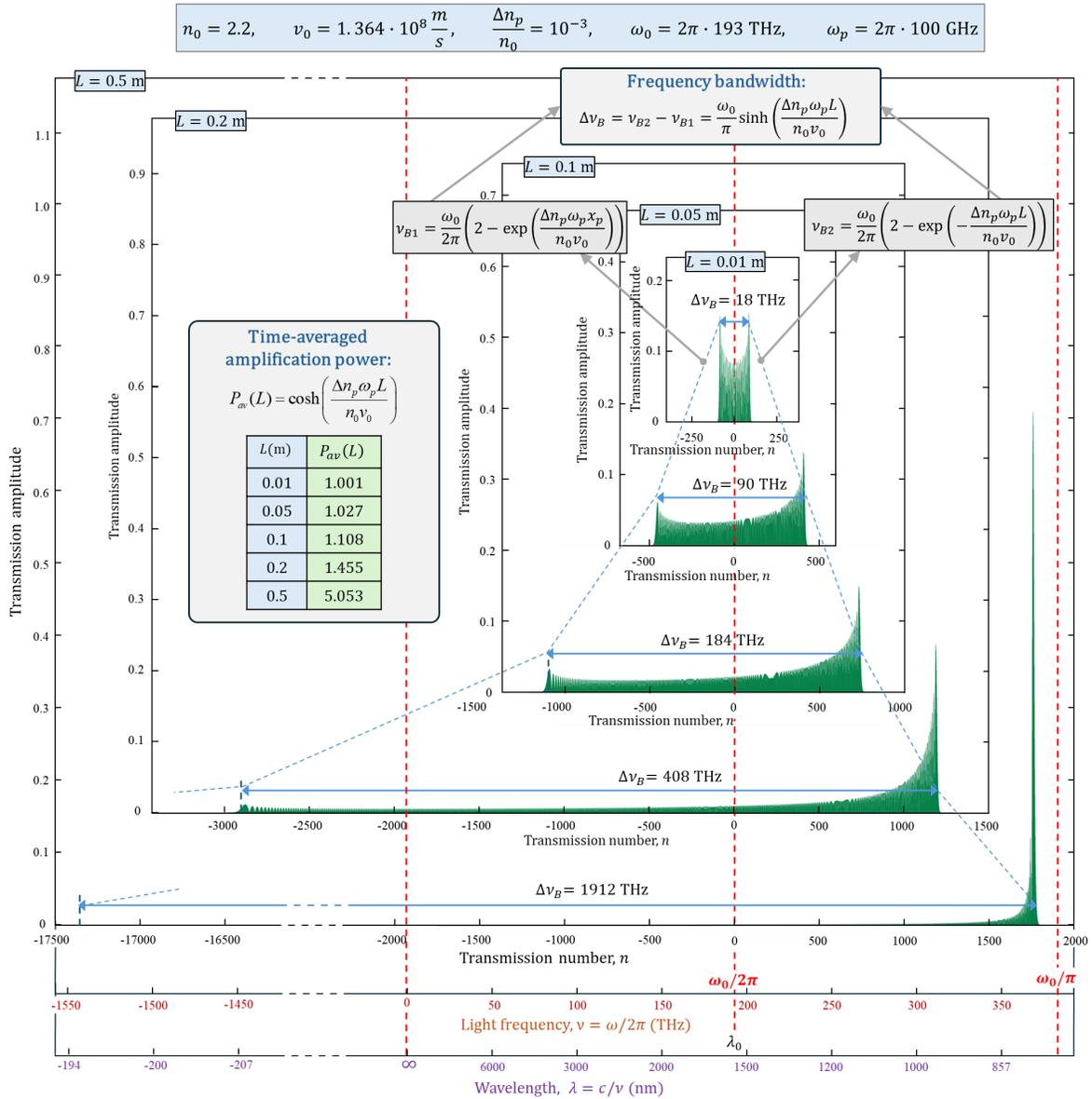

**Fig. 4.** Transmission spectra and time-average amplification power at the completely synchronous condition $v_p = v$ for different modulation lengths $L = 0.01, 0.05, 0.1, 0.2,$ and $0.5$ m. Parameters of the input light, waveguide, and modulation are indicated at the top of the figure.

The feasibility of amplification of light by a synchronous traveling wave can be better understood by comparing Eqs. (12) and (17). From these equations, we find a simple relation between the time-averaged normalized power and the spectral bandwidth of the outgoing wave:

$$P_{av}(L) = \sqrt{1 + \left(\frac{\Delta\omega_B(L)}{2\omega_0}\right)^2}. \tag{18}$$

Thus, significant time-averaged amplification is impossible if the spectrum bandwidth $\Delta\omega_B(L)$ is small compared to the input frequency $\omega_0$. Eq. (18) is derived under the condition of dispersionless propagation, which, for realistic optical waveguides, can be valid only within a relatively small bandwidth $\Delta\omega_B$. Since the transmission bandwidths of optical materials are also relatively small, the significant amplification of light in optical waveguides parametrically modulated by a traveling wave is currently unfeasible.

### E. Wave propagation and wave spectrum in a completely asynchronous case $|\mu| \ll 1$

In contrast to the amplification, the traveling wave modulation with the phase velocity $v_p$ approaching the phase velocity of light $v_0$ is important for the enhancing the performance of broadband optical modulators and frequency comb generators [23-30, 55, 56]. The frequency comb bandwidth, which is close to maximum possible for a given modulation length $L$ and amplitude $\Delta n_p$, can be achieved without the accurate proximity to the completely synchronous condition $v_p = v_0$ (i.e., satisfying the condition $|\mu| \gg 1$ and even $|\mu| > 1$) considered in the previous section. Here we demonstrate these results considering the asynchronous case

$$|\mu| = \frac{\Delta n_p v_p}{n_0 |v_p - v_0|} \ll 1. \tag{19}$$

Keeping the zero and the first order in $|\mu|$ and $\eta_0$ terms in the expression for the wave amplitude $U_0(x,t)$ and phase $S(x,t)$ given by Eqs. (5)-(8), we find:

$$E(x,t) = U_0(x,t)\exp(iS(x,t)),$$

$$S(x,t) = \omega_0\left(\frac{x}{v_0} - t\right) + i\frac{\omega_0 \eta_0}{c}x + \Omega_p(x)\cos\left(\omega_p\left(t - \frac{(v_0 + v_p)}{2v_0 v_p}x\right)\right), \tag{20}$$

$$U_0(x,t) = 1 - \Omega_p(x)\frac{\omega_p}{2\omega_0}\left(\frac{v_0}{v_p} - 3\right)\sin\left(\omega_p\left(t - \frac{(v_0 + v_p)}{2v_0 v_p}x\right)\right).$$

In this equation, we introduce the *modulation index* $\Omega_p(x)$ which characterizes the effect of modulation on the propagation of a wave along the modulation length $L$:

$$\Omega_p(L) = \frac{2\Delta n_p \omega_0 v_p}{n_0 \omega_p (v_0 - v_p)}\sin\left(\frac{(v_0 - v_p)}{2v_0 v_p}\omega_p L\right). \tag{21}$$

The structure of the eikonal $S(x,t)$ in Eq. (20) resembles the expressions known in the theory of optical modulators [25, 28]. In particular, it follows from this equation that, in this case, the effect of material losses is described by the factor $\exp(-\omega_0 \eta_0 x/c)$, which is the same as for the stationary (unmodulated) wave propagation. For the case of instantaneous modulation, $v_p = \infty$, Eq. (21) coincides with that for the absolute value of modulation index found in Ref. [57]. The expression for the normalized amplitude $U_0(x,t)$ in Eq. (20) shows that its deviations from unity is commonly small due to the factor $\omega_p/\omega_0 \ll 1$. It also follows from this expression that the first order in modulation amplitude $\Delta n_p$ term in the time-averaged power $P_{av}(x)$ defined by Eq. (9) vanishes. Taking into account the second order in $\Delta n_p$ (more precisely – in $\mu$) term, we find:

$$P_{av}(L) \cong 1 - \frac{2\omega_0 \eta_0}{c} x + \frac{(v_0 - 2v_p)(v_0 - 3v_p)}{v_p^2}\left(\Omega_p(L)\frac{\omega_p}{\omega_0}\right)^2 \quad (22)$$

$$= 1 - \frac{2\omega_0 \eta_0}{c} L + \frac{\Delta n_0^2}{4n_0^2}\frac{(v_0 - 2v_p)(v_0 - 3v_p)}{(v_0 - v_p)^2}\sin^2\left(\frac{v_0 - v_p}{2v_0 v_p}\omega_p L\right).$$

This equation shows that, close to the completely synchronous condition $v_0 = v_p$ or, more precisely, for $|v_0 - v_p| \ll v_0$ and modulation lengths satisfying the inequality

$$L \ll L_s = \frac{2v_0 v_p}{\omega_p |v_0 - v_p|}, \quad (23)$$

the modulation leads to a small wave amplification equal to

$$P_{av}(L)\Big|_{L \ll L_s} \cong 1 - \frac{2\omega_0 \eta_0}{c} L + \frac{\Delta n_0^2 \omega_p^2}{2c^2} L^2. \quad (24)$$

Unexpectedly, for zero losses, $\eta_0 = 0$, this result coincides with that given by Eq. (12) for $\Delta n_0 \omega_p L/c \ll 1$. This result is also similar to that found in Ref. [17] under the same assumption $|\mu| \ll 1$. From Eq. (24), the modulation length leading to the noticeable amplification of light can be estimated as $L_a = c/(\Delta n_0 \omega_p)$. For the parameters considered here, $\omega_p \sim 2\pi \cdot 100$ GHz, $n_0 = 2.2$, $\Delta n_0/n_0 \sim 10^{-3}$, we have $L_a \sim 20$ cm. Eq. (22) shows that $P_{av}(L)$ vanishes if the relation between phase velocities is close to $v_0 = 2v_p$ and $v_0 = 3v_p$. Alternatively, modulation leads to the wave attenuation if $2v_p < v_0 < 3v_p$. It follows from Eq. (22) that amplification is always small outside the vicinity where $|v_0 - v_p|/v_p \ll 1$. This result is also evident from the solutions given by Eq. (5)-(8) for $|\mu| \ll 1$ and $\Delta n_p/n_0 \ll 1$. Eq. (22) will be used below to discuss the relation between the possible amplification and the required transmission bandwidth.

The characteristic dependencies of $\Omega_p(L)$ as a function of the ratio $v_p/v$ at different modulation lengths, $L = 0.05$ mm, $0.68$ mm, $4.77$ mm, and $49.8$ mm, are shown in Fig. 5. In this figure, we again assume that the waveguide refractive index is that of lithium niobate, $n_0 = 2.2$, the relative amplitude of refractive index modulation is $\Delta n_0/n_0 = 10^{-3}$, and the light and modulation frequencies are $\omega_0 = 2\pi \cdot 193$ THz and $\omega_p = 2\pi \cdot 100$ GHz. Due to the periodic dependence of $\Omega_p(L)$, the values of $L$ in Fig. 5 correspond to the maxima of $\Omega_p(L)$ nearest to the lengths $L = 0.05, 0.5, 5,$ and $50$ mm at $v_p = \infty$. From Eq. (21), these maxima are situated at the periodic sequence of modulation lengths $L = \frac{\pi v_0}{\omega_p}(2N + 1)$, $N = 0, 1, 2 \ldots$. It is seen from the plots of Fig. 5 that, while the modulation index for the completely synchronous modulation ($v = v_p$) and instantaneous modulation ($v_p = \infty$) are close to each other for the small modulation length

$L \lesssim 0.1$ mm, the modulation index becomes much greater for larger modulation lengths at $v_p$ approaching $v$. Due to the condition of Eq. (19), the synchronous case and, in particular, the exact equality $v = v_p$ is excluded in the considered approximation. However, as follows from Eq. (19), the plots in Fig. 5 are accurate everywhere except for a relatively small vicinity of the completely synchronous coordinate $v_p/v = 1$ where $|v_p/v - 1| \sim \Delta n_p/n_0 = 10^{-3}$. Therefore, these plots are reasonably accurate for all $v_p/v$. Assuming that the modulation length $L$ is sufficiently small as defined by Eq. (23), or that $v_p/v \to 1$, we simplify Eq. (21) for $\Omega_p(x)$ to

$$\Omega_p(L)\Big|_{L \ll L_s} = \frac{\Delta n_p \omega_0}{c} L. \qquad (25)$$

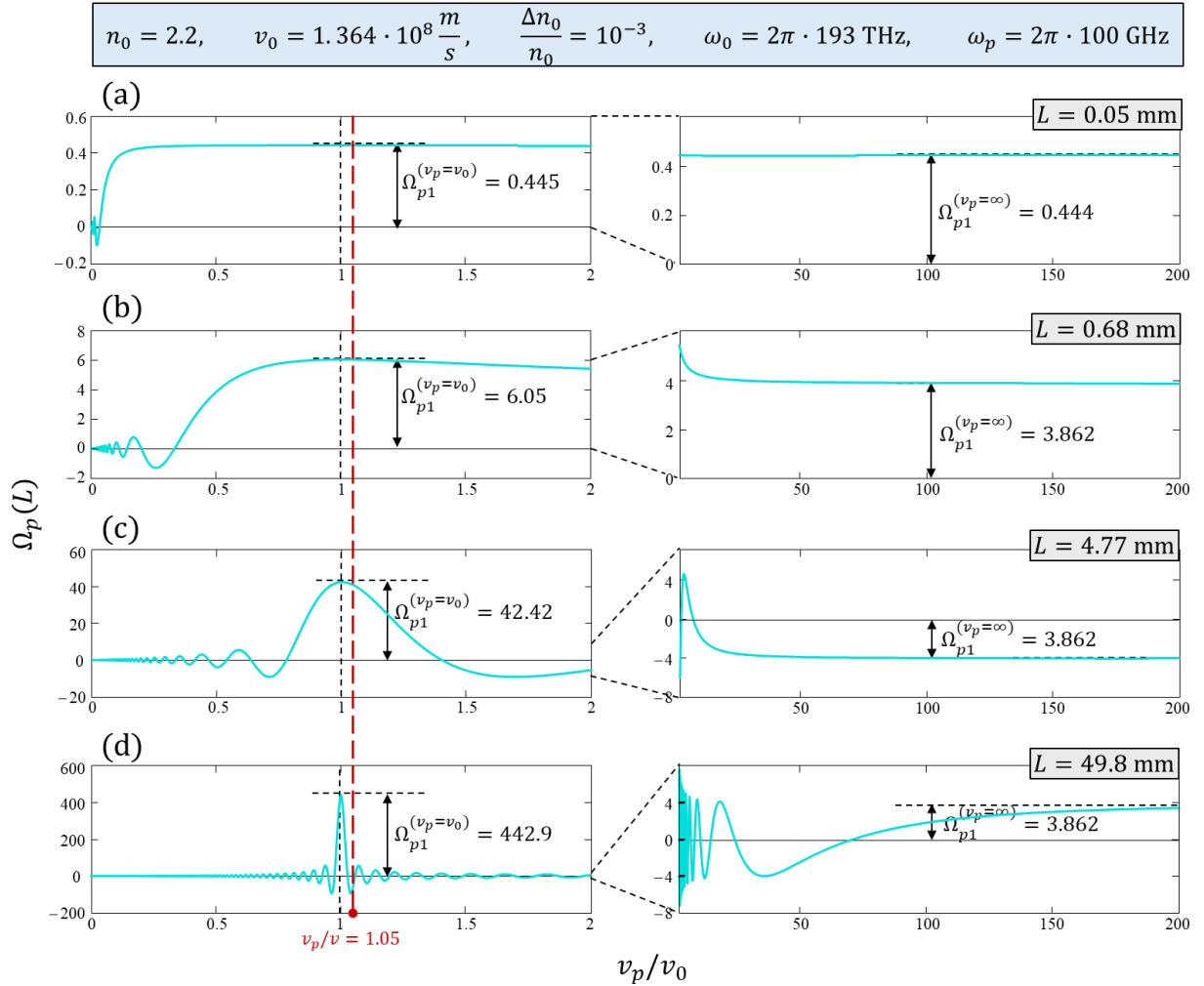

**Fig. 5.** Modulation index $\Omega_p$ as a function of the ratio $v_p/v_0$ at different modulation lengths, (a) $L = 0.05$ mm, (b) $L = 0.68$ mm, (c) $L = 4.77$ mm, and (d) $L = 49.8$ mm, for the lithium niobate waveguide. The light and modulation parameters are shown at the top of the figure.

From this equation, the maximum of modulation index at $v \cong v_p$ grows linearly with the modulation length $L$. Indeed, Fig. 5 shows that the modulation index can be dramatically increased for large $L$ if the phase velocity of the traveling wave is sufficiently close to the phase velocity of light.

While, for realistic waveguides, the substantial amplification of light is unfeasible, the modulation index $\Omega_p(L)$ defined by Eq. (21) can significantly exceed unity and lead to the creation of a relatively broadband comb spectrum near the synchronous condition $v_p = v$ (see Fig. 5(c) and (d)). To determine the spectrum of the output wave for $|\mu| \ll 1$, we rewrite Eq. (20) as

$$E(x,t) = \exp\left[i\omega_0\left(\frac{x}{v_0} - t\right) - \frac{\omega_0 \eta_0}{c}x + i\Omega_p(x)\cos\left(\omega_p\left(t - \frac{v_0 + v_p}{2v_0 v_p}x\right) - iG_p\right)\right]. \tag{26}$$

Here we introduce a small parameter

$$G_p = \frac{\omega_p}{2\omega_0}\left(\frac{v_0}{v_p} - 3\right), \quad |G_p| \ll 1. \tag{27}$$

We note that $|G_p| \ll 1$ due to Eq. (4). Applying the Jacobi-Unger expansion to Eq. (26) we find:

$$E(L,t) = \sum_{m=-\infty}^{\infty} U_m^{(c)} \exp\left[i(m\omega_p - \omega_0)t\right],$$

$$U_m^{(c)} = J_m(\Omega_p(L))\exp\left[-\frac{\omega_0 \eta_0}{v}x_p + \frac{i\pi m}{2} + i\omega_0\frac{L}{v} - i\omega_p\frac{v_0 + v_p}{2v_0 v_p}Lm + G_p m\right]. \tag{28}$$

To estimate the maximum possible amplitude of conversion $\omega_0 \to \omega_0 + |m|\omega_p$, we note that, while the maximum argument of the Bessel function in Eq. (28) can be large, $z = \Omega_p(L) \gg 1$, the maximum of $|J_m(z)|$ is always smaller than $2^{-1/2}$ [58]. For $|m| \gg 1$, the $|J_m(z)|$ maximum is defined by its asymptotics equal to $0.674|m|^{-1/3}$ [58], i.e., vanishes very slowly. This maximum is achieved at $z \cong |m|$, while $|J_m(z)|$ rapidly vanishes for $|z| > |m|$. Thus, the frequency comb bandwidth of $E(x_p, t)$ is determined from Eq. (28) as

$$\Delta\omega_B(L) = 2\omega_p|\Omega_p(L)|, \tag{29}$$

and the amplification of a comb line is determined from the Eqs. (27) and (28) by the factor

$$F_m = \exp(mG_p) = \exp\left[m\frac{\omega_p}{2\omega_0}\left(\frac{v_0}{v_p} - 3\right)\right]. \tag{30}$$

Since in the approximation considered $|G_p| \ll 1$, the factor $F_m$ can be large only for sufficiently large positive comb line numbers $m$. From Eqs. (29) and (30), we find the maximum possible amplification takes place for the maximum positive $m = |\Omega_p(L)|$ within this bandwidth. Then, we find from Eq. (21) for $\Omega_p(L)$ that the amplification effect defined by $F_m$ is always small away from the synchronous condition, confirming the general result directly following from Eq. (22).

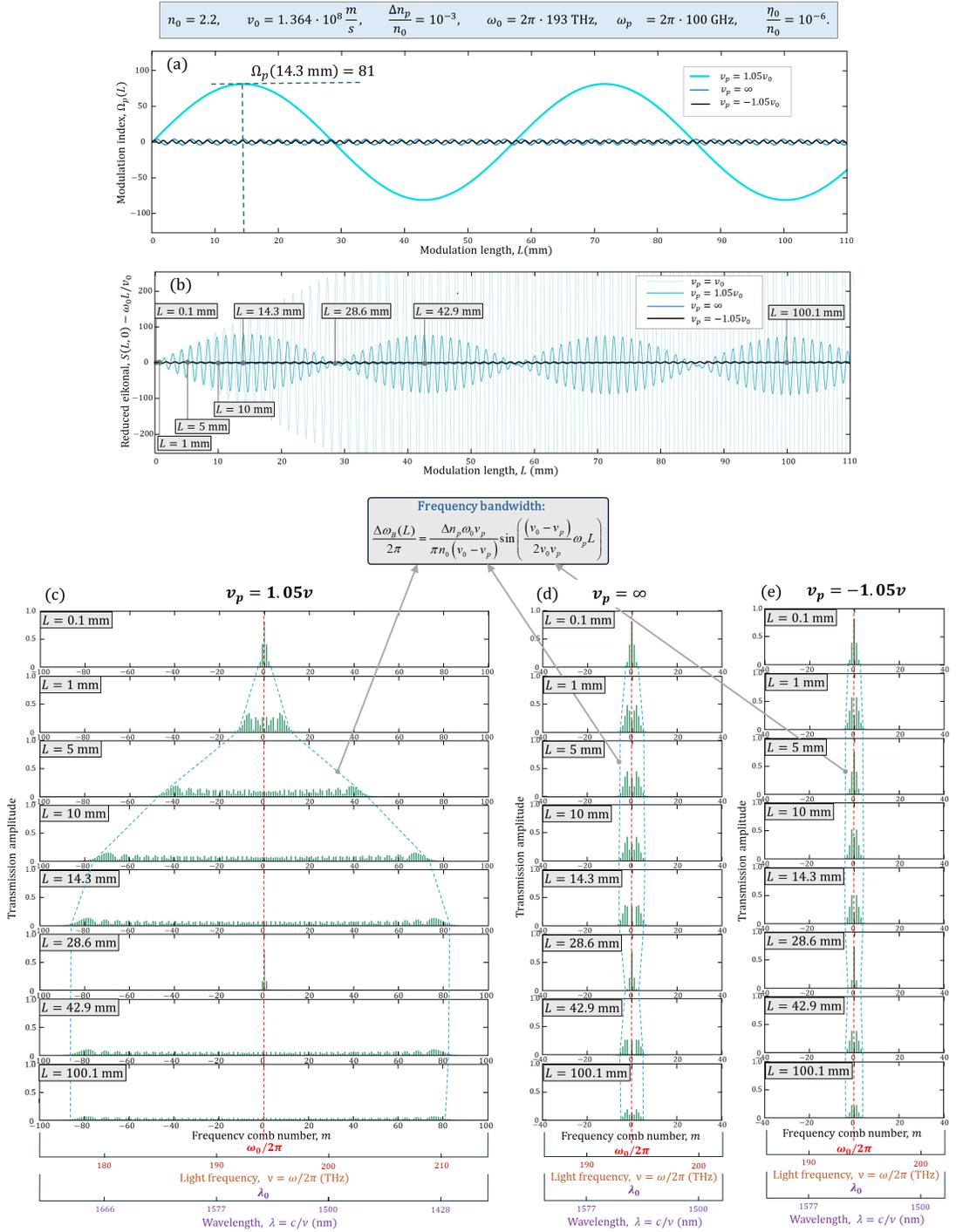

**Fig. 6.** (a) Dependences of modulation index $\Omega_p(L)$ on the modulation length $L$ for close phase velocities, $v_p = 1.05v$, in the asynchronous case $|\mu| = 0.02$ (light blue curve), for the instantaneous modulation, $v_p = \infty$ (blue curve), and for the reverse modulation $v_p = -1.05v$ (black curve). (b) Dependences of the reduced eikonal, $S(L,0) - \omega_0 L/v_0$, on the modulation length $L$ for the completely synchronous modulation, $v_p = v$ (dimmed light blue curve), $v_p = 1.05v$ (light blue curve), $v_p = \infty$ (blue curve), and $v_p = -1.05v$ (black curve). (c), (d), and (e) The transmission amplitudes for (c) $v_p = 1.05v$, (d) $v_p = \infty$, and (d) $v_p = -1.05v$ for different modulation lengths $L$ indicated in the plots of this figure.

As noted above, proximity to the completely synchronous condition $v_p = v$ can significantly increase the modulation index (up the values of ~ 40 and ~ 400 for the lithium niobate waveguide with $L \cong 5$ mm and $L \cong 50$ mm, respectively, see Fig. 5) and as follows from Eq. (29) can increase the frequency comb bandwidth $\Delta\omega_p$ proportionally. A dramatic enhancement of the spectral bandwidth generated by a traveling wave having the phase velocity close to $v$ though still for $|\mu| \ll 1$, as compared to the bandwidth generated by the instantaneous modulation with $v_p = \infty$ and reverse modulation with the reverse sign of $v_p$, is evidenced from Fig. 6. The reason for the enhancement is a much greater value of the modulation index of a traveling wave having $v_p \cong v_0$. The parameters of light, waveguide, and modulation, which are similar to the parameters considered in our previous examples, are indicated at the top of this figure. For the waveguide with these parameters, the value of $\Omega_p(L)$ achieves 81 at $L = 14.3$ mm (Fig. 6(a)) at $v_p = 1.05v$, though remains much smaller for the instantaneous and reverse modulations when $v_p = \infty$ and $v_p = -1.05v_0$, respectively. Figs. 6(b) shows the dependences of the reduced eikonal, $S(x_p, 0) - \omega_0 x_p/v_0$, on the modulation length for the same phase velocity relations as well as for the completely synchronous case $v_p = v_0$. It is seen that the oscillation amplitude of the eikonal as a function of modulation length is proportional to the local modulation index, as directly follows from Eq. (21). Figs. 6(c), (d), and (e) compare the frequency comb spectra of solutions for (c) $v_p = 1.05v_0$, (d) $v_p = \infty$, and (e) $v_p = -1.05v$ for different modulation lengths $L$ indicated in the plot (b) of this figure. It is seen that, in accordance with Eq. (29), the generated frequency comb bandwidth is proportional to the value of corresponding modulation index shown in Fig. 6(a). We note that the dramatic reduction of modulation effect of the reverse vs. the direct modulation manifests the strong nonreciprocity of the considered device.

In contrast to the optical frequency comb bandwidth, the total amplification of light induced by modulation remains small for $\Delta\omega_B(L) \ll \omega_0$. Indeed, comparing Eq. (29) and Eq. (22), we find:

$$P_{av}(L) \cong 1 - \frac{2\omega_0 \eta_0}{c} L + \frac{(v_0 - 2v_p)(v_0 - 3v_p)}{v_p^2}\left(\frac{\Delta\omega_B(L)}{4\omega_0}\right)^2. \tag{31}$$

In particular, close to the completely synchronous condition $v_p = v_0$

$$P_{av}(L)\Big|_{v_p \to v_0} = 1 - \frac{2\omega\eta_0}{c}L + \frac{1}{8}\left(\frac{\Delta\omega_B(L)}{\omega_0}\right)^2. \tag{32}$$

Remarkably, this equation coincides with Eq. (18) for relatively small bandwidth $\Delta\omega_B \ll \omega_0$. Assuming that velocities $v_0$ and $v_p$ are of the same order, we find from Eq. (31) that similar to the completely synchronous case described by Eq. (18), significant amplification of light is impossible in realistic waveguides, which always have $\Delta\omega_B \ll \omega_0$. However, the situation for small traveling wave velocities $v_p \ll v_0$ cannot be clarified from Eq. (31) due to the restriction of the eikonal approximation, $v_p \gg v_0 \omega_p/\omega_0$ following from Eq. (4). The latter restriction will be removed in the following Section.

### III. THE PERTURBATION THEORY APPROACH

The eikonal approximation used above does not allow us to consider sufficiently small values of $v_p$ since, according to Eq. (4), the condition of slowness of modulation in space restricts these values to $v_p \gg \omega_p v_0/\omega_0$. However, we will show below in Section IV that the case of comparable $\omega_p/v_p \sim \omega_0/v_0$ is important to arrive at the strong amplification of light in an optical resonator. We find from Eq. (21) that for $\omega_p/v_p \sim \omega_0/v_0$ and $\omega_p \ll \omega_0$ the modulation index $|\Omega_p(L)| \sim \Delta n_p/n_0 \ll 1$. Under the latter condition, the restriction $\omega_p/v_p \ll \omega_0/v_0$ can be withdrawn and solution of the wave equation, Eq. (1), can be found by the regular perturbation theory.

Having in mind modulation by acoustic traveling waves, which strongly attenuate in space [59, 60], we consider now the refractive index in Eq. (1) in the form (Fig. 7):

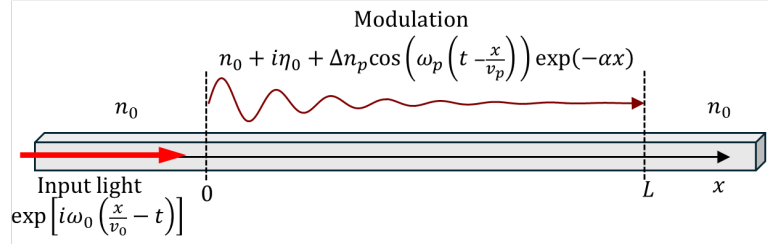

**Fig. 7.** An optical waveguide with the refractive index modulated by a traveling wave fully attenuating within the interval $(0, L)$.

$$n(x,t) = n_0 + \Delta n(x,t),$$

$$\Delta n(x,t) = i\eta_0 + \Delta n_p \cos\left(\omega_p\left(t - \frac{x}{v_p}\right)\right)\exp(-\alpha x) \tag{33}$$

Here $\alpha$ defines the attenuation of modulating wave. In the first order over $\Delta n_p/n_0$ and $\eta_0/n_0$, the solution of Eq. (1) with the boundary condition

$$E^{(0)}(x,t) = \exp\left(i\omega_0\left(\frac{x}{v_0} - t\right)\right) \tag{34}$$

is found as

$$E(x,t) = E^{(0)}(x,t)\left(1 + \Delta U_p(x,t)\right), \tag{35}$$

where $|\Delta U_p(x,t)| \ll 1$. Under the latter condition, it is convenient to present this solution in the form similar to that in Eq. (26) using $1 + \Delta U_p(x,t) \cong \exp(\Delta U_p(x,t))$. Then calculations detailed in Appendix D yield:

$$E(x,t) = E^{(0)}(x,t)\exp\left(-\eta_0\frac{\omega_0 x}{c} + i\tilde{\Omega}_p(x)\cos\left(\omega_p\left(t - \frac{v_0 + v_p}{2v_0 v_p}x\right) - i\tilde{G}_p(x)\right)\right), \tag{36}$$

$$\tilde{\Omega}_p(x) = -2i\sqrt{\Delta U^+ \Delta U^- W^+(x)W^-(x)}, \quad \Theta(x) = \frac{1}{2}\ln\left(\frac{\Delta U^- W^-(x)}{\Delta U^+ W^+(x)}\right), \tag{37}$$

where

$$W^\pm(x) = \sin\left(\pm\frac{v_0 - v_p}{2v_0 v_p}\omega_p x + \frac{i\alpha}{2}x\right)\exp\left(-\frac{\alpha}{2}x\right), \tag{38}$$

$$\Delta U^{\pm} = \frac{\Delta n_p v_p^2 (\omega_0 \pm \omega_p)^2}{n_0 \left[\pm \omega_p (v_0 - v_p) + i\alpha v_0 v_p\right]\left[2\omega_0 v_p \pm \omega_p (v_0 + v_p) + i\alpha v_0 v_p\right]}. \tag{39}$$

This solution is valid only if $|\Delta U_p(x,t)| \ll 1$, or

$$\left|\Delta U^{\pm}\right| \ll 1. \tag{40}$$

Obviously, in contrast to the eikonal approximation, solution described by Eqs. (36)-(39) includes transitions with acquisition or loss of a single frequency $\omega_p$ only. For convenience, this solution is chosen so that it vanishes at the starting point of modulation, $x = 0$, $\Delta U_p(0,t) = 0$. At zero attenuation, $\alpha = 0$, under the conditions of Eq. (40) and Eq. (4) this solution coincides with that given by Eq. (26). In particular, $\widetilde{\Omega}_p(x)$ coincides with $\Omega_p(x)$ and $\tilde{G}_p(x)$ coincides with $G_p(x)$.

Similar to the eikonal approximation, for the zero attenuation of modulation, $\alpha = 0$, the denominator in the expressions for $\Delta U^{\pm}$ vanishes at the synchronous condition $v_p = v_0$, however, this does not lead to the singularity of determined solution. In contrast, for $\alpha = 0$ this solution *possesses a singularity* at

$$2\omega_0 v_p = \pm \omega_p (v_0 + v_p), \tag{41}$$

where $\Delta U^{\pm}$ can significantly grow, yet under the restriction of Eq. (40). The sign $\pm$ in Eq. (41) corresponds to the co- and contra-propagating light and traveling waves. Naturally, Eq. (41) coincides with the phase matching condition for the backward Brillouin scattering [47]. For the case of our concern, $v_p \ll v_0$ and $\omega_p \ll \omega_0$, this singularity – missed in eikonal approximation which requires slowness of modulation in space, $k_p = \omega_p/v_p \ll k_0 = \omega_0/v_0$ (see Eq. (4)) – takes place close to the condition when the propagation constant $k_p$ of the traveling wave is twice as large as the propagation constant $k_0$ of light, i.e., at $\omega_p/v_p = 2\omega_0/v_0$, similar to the Brillouin scattering [47].

For determinacy, the resonance of $\Delta U^{-}$ corresponding to sign + in Eq. (41) will be considered below when the modulation frequency

$$\omega_p^{(res)} = \frac{2\omega_0 v_p}{v_0 + v_p}. \tag{42}$$

At this resonance, under the conditions of our interest $v_p \ll v_0$ and $\omega_p \ll \omega_0$ and commonly valid condition $\alpha v_p \ll \omega_p$ (see below) we have

$$\Delta U^{-} = \Delta U^{-}_{res} = \frac{i\Delta n_p \omega_0}{2\alpha n_0 v_0}. \tag{43}$$

Close to this resonance, the part of solution including $\Delta U^{+}$ can be neglected. To estimate the maximum possible amplification effect, we assume that the propagation length $x$ is sufficiently large so that $\exp(-\alpha x) \ll 1$. Then, solution presented by Eqs. (36)-(39) is simplified to

$$E(x,t) = E^{(0)}(x,t)\left[1 + \Delta U^{-}_{res} \exp\left(i\omega_p \left(t - \frac{x}{v_0}\right)\right)\right]. \tag{44}$$

The maximum averaged over time amplification of light propagating along the waveguide with the length exceeding $\alpha^{-1}$ is determined as (see Appendix D)

$$P_{av} \cong 1 + \left(\frac{\Delta n_p \omega_0}{2\alpha n_0 v_0}\right)^2. \tag{45}$$

As an example, for a lithium niobate waveguide, we set $v_p \sim 4000$ m/s, which yields $\omega_p \cong \frac{2\omega_0 v_p}{v_0} \sim 2\pi \cdot 10$ GHz. Setting $\Delta n_p \cong 2 \cdot 10^{-4}$ and the smallest demonstrated attenuation $\alpha \cong 500$ m$^{-1}$ at this modulation frequency in the bulk lithium niobate [59], we have $P_{av} - 1 \cong 0.02$. For a larger initial index modulation $\Delta n_p \sim 10^3$, which is much more challenging to achieve, the amplification may become comparable with the original light power and the perturbation approach fails.

## IV. TRANSFORMATION AND AMPLIFICATION OF LIGHT BY AN OPTICAL RESONATOR

We consider now a closed optical waveguide with length $2L$ forming a racetrack resonator which is coupled to an input-output waveguide as illustrated in Fig. 8. We assume that the modulation is described by Eq. (1) with the refractive index defined by Eq. (2) (Section IV.A) and by Eq. (33) (Section IV.B) and takes place along the length $L$ of the resonator waveguide. The monochromatic input light in the input-output waveguide near position $x = x_0$ in front of the coupling region is set to $E_{in}(x,t) = \exp(i\omega_0(x/v_0 - t))$. Similar to the previous Section II.E, we consider the asynchronous modulation with $|\mu| \ll 1$. Using the transfer matrix approach (see, e.g., [61]), we find the output light field $E_{out}(t)$ from the equation

$$\begin{pmatrix} E_{out}(t) \\ E(x_0,t) \end{pmatrix} = S \begin{pmatrix} E_{in}(t) \\ E(2L+x_0,t) \end{pmatrix}, \quad S = \begin{pmatrix} \tau & \kappa \\ -\kappa & \tau \end{pmatrix}, \tag{46}$$

where matrix $S$ is the unitary S-matrix, so that $\tau^2 + \kappa^2 = 1$. In this equation, the coordinates $x = 2L + x_0$ and $x = x_0$ define the beginning and the end of the coupling region and the S-matrix parameters $\kappa$ and $\tau$ determine the coupling between the input-output waveguide and resonator (Fig. 8).

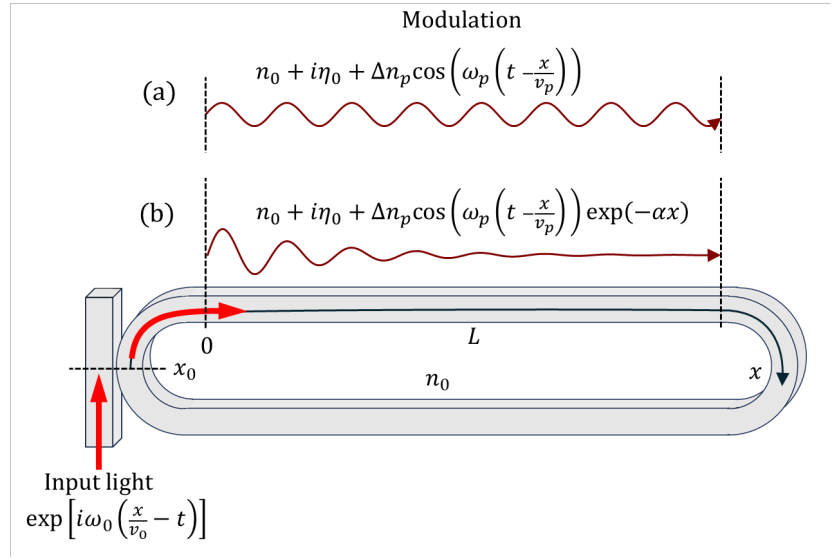

Fig. 8. An optical racetrack resonator with the refractive index modulated by a traveling wave along the interval $(0, L)$. (a) Uniform modulation. (b) Attenuating modulation.

## A. The eikonal approximation

In our calculations, we follow the approach of Ref. [57] where the determination of the output wave $E_{out}(t)$ was reduced to the solution of a functional equation. For modulation without attenuation (Fig. 8(a)), using Eq. (24) we find:

$$E(x_0,t) = \exp(-i\omega t)\Phi(t), \tag{47}$$

$$E(2L+x_0,t) = \exp(-i\omega t) A(t)\Phi(t-T), \quad T = \frac{2L}{v_0}, \tag{48}$$

$$A(t) = \exp\left[i\omega_0 T - \frac{\eta_0}{n_0}\omega_0 T + i\Omega_p(L)\cos\left(\omega_p\left(t - \frac{(v_0+v_p)}{2v_0 v_p}L - iG_p\right)\right)\right]. \tag{49}$$

Here $T$ is the roundtrip circulation time. Eqs. (47)-(49) lead to the functional equation for the arbitrary function $\Phi(t)$:

$$\Phi(t) = \tau A(t)\Phi(t-T) - \kappa \tag{50}$$

which can be solved exactly [57]. As the result, in full analogy with calculations of Ref. [57] (see Appendices B and C of [57]), the comb spectral amplitudes $U_m^{(c)}$ of $E_{out}(t)$ are found from the expansion:

$$E_{out}(t) = \sum_{m=-\infty}^{\infty} U_m^{(c)} \exp\left[i(m\omega_p - \omega_0)t\right],$$

$$U_m^{(c)} = \tau\delta_{0m} - \kappa^2 \exp\left[im\left(\frac{\pi}{2} - \frac{\omega_p T}{2} - \frac{\omega_p(v_0+v_p)}{2vv_p}L - iG_p\right)\right] \times$$

$$\sum_{n=0}^{\infty} \tau^n J_m\left(\sigma_{n+1}\Omega_p(L)\right)\exp\left[(n+1)\left(-\frac{im}{2}\omega_p T + i\omega_0 T - \frac{\eta_0}{n_0}\omega_0 T\right)\right], \tag{51}$$

$$\sigma_n = \frac{\sin\left(\frac{n}{2}\omega_p T\right)}{\sin\left(\frac{1}{2}\omega_p T\right)}, \quad G_p = \frac{\omega_p}{2\omega_0}\left(\frac{v_0}{v_p} - 3\right),$$

where $\delta_{nm}$ is the Kronecker delta and the amplification parameter $G_p$ is the same as in Eq. (27). The total time-averaged output power is calculated from this equation as

$$P_{av}^{(out)} = \sum_{m=-\infty}^{\infty} \left|U_m^{(c)}\right|^2. \tag{52}$$

In our further calculations we assume that coupling between the input-output and resonator waveguides $\kappa$ is small so that $\tau \cong 1 - \kappa^2/2$ and, in Eq. (51), $\tau^n \cong \exp(-n\kappa^2/2)$. Under this assumption, the microresonator Q-factor found from Eq. (51) is

$$Q = \frac{1}{2}\left(\frac{\eta_0}{n_0} + \frac{\kappa^2}{2\omega_0 T}\right)^{-1} \qquad (53)$$

and its intrinsic Q-factor is $Q_{int} = n_0/(2\eta_0)$.

Similar to Eq. (28), which describes the nonresonant propagation, the difference of the expressions for the spectral amplitudes $U_m^{(c)}$ determined here for the traveling wave modulation compared to those previously found for the instantaneous modulation [57], consists in a different expression for modulation index $\Omega_p(L)$ defined now by Eq. (21), the additional phase factor $\exp(-im\omega_p(v_p + v_0)L/(2v_0v_p))$, and amplification factor $F_m$ defined by Eq. (30). Now, in contrast to the nonresonant propagation, the factor $F_m$, which can be large for a sufficiently small traveling wave phase velocity $v_p$ and large comb line numbers $m$. Therefore, this factor can significantly increase the total output power as well as the power of individual frequency comb lines.

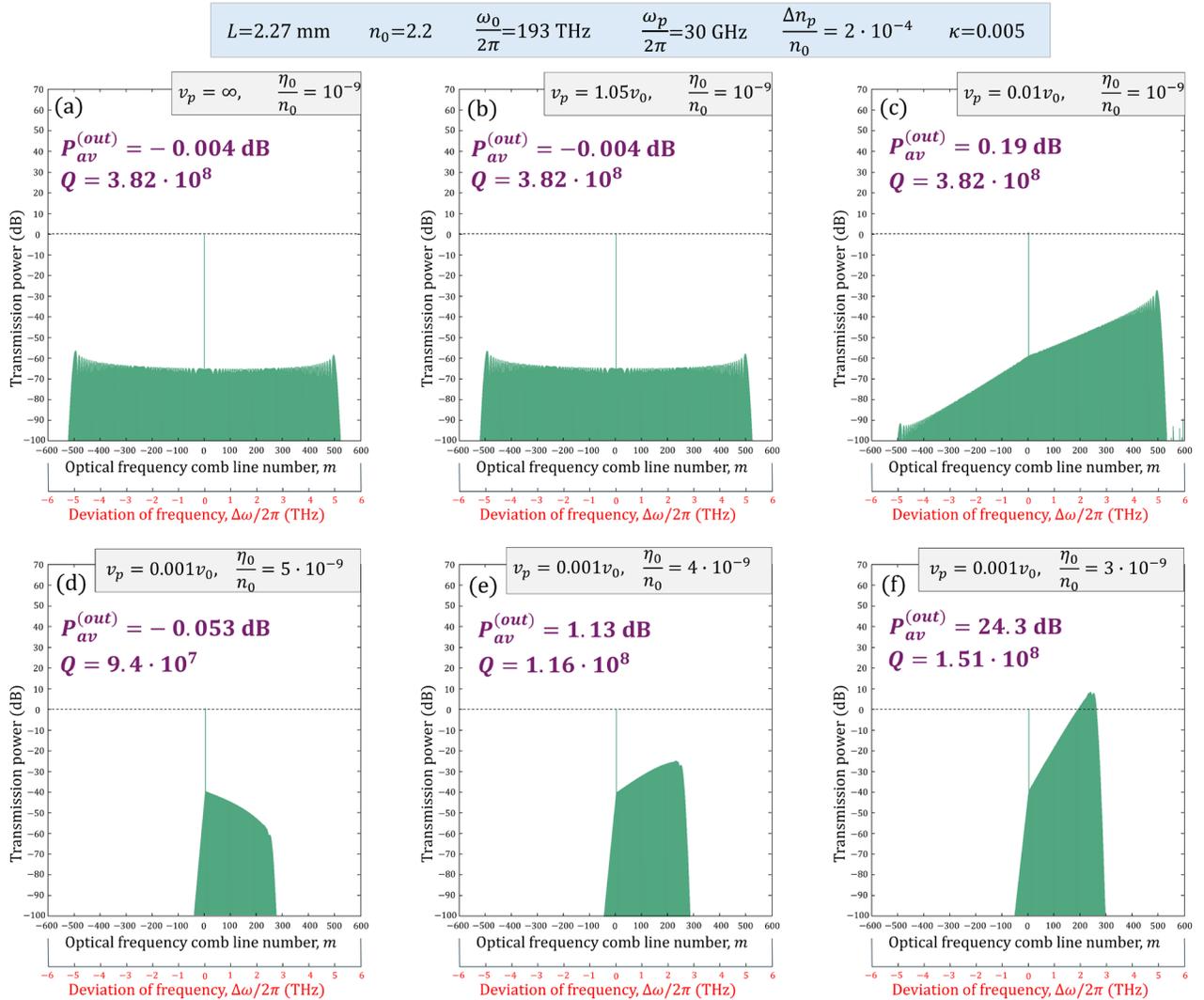

**Fig. 9.** The resonant transmission power spectra for a racetrack resonator. The system parameters are indicated on the top of the figure. Plots (a)-(f) correspond to different traveling wave velocities $v_p$ and waveguide propagation losses $\eta_0$. These and other parameters of the plots are detailed in Table 1.

The maximum amplitude and bandwidth of the output wave spectrum defined by Eq. (51) is achieved at the exact optical and modulation resonances, determined, respectively, by equations $\omega_0 T = 2\pi N_0$ and $\omega_p T = 2\pi N_p$ with integers $N_0 \gg 1$ and $N_p$. It becomes clear from the expression for $U_m^{(c)}$ in Eq. (51) that the deviation of the modulation frequency from this resonance condition,

$$\Delta\omega_p = \omega_p - \frac{2\pi N_p}{T}, \tag{54}$$

will reduce the magnitude of $U_m^{(c)}$. The reduction, growing with the frequency comb number $m$, is due to the term $-i(n+1)m\Delta\omega_p T/2$ in the exponent of the sum over $n$ (rewritten using Eq. (54)). Consequently, choosing the appropriate offset $\Delta\omega_p$, we can control the transmission bandwidth $\Delta\omega_B$ of the resonator. This result, illustrated in Fig. 5(b) of Ref. [62] for the instantaneous modulation ($v_p = \infty$), suggests that choosing an appropriate offset $\Delta\omega_p$ we can appropriately shrink the transmission bandwidth $\Delta\omega_B$ and simultaneously increase (rather than preserve or attenuate) the transmission power within this bandwidth. We show below that choosing a transmission band with a relatively small width (which, in practice, should correspond to the band with the smallest material loss and dispersion) we can also redirect and amplify light within this band only.

| Resonator half-length $L$ (mm) | Refractive index $n_0$ | Light frequency $\frac{\omega_0}{2\pi}$ (GHz) | Refractive index modulation $\Delta n_p$ | Modulation frequency $\frac{\omega_p}{2\pi}$ (GHz) | Coupling coefficient $\kappa$ |
|---|---|---|---|---|---|
| 2.27 | 2.2 | 193,000 | $4.4 \cdot 10^{-4}$ | 30 | 0.005 |

| Plot | Travelling wave phase velocity $v_p$ | Relative material loss $\eta_0/n_0$ | Material loss $20\log(e) \times \eta_0\omega_0/c$ (dB/m) | Modulation index $\Omega_p$ | Deviation from modulation resonance $\frac{\Delta\omega_p}{2\pi}$ (MHz) | Intrinsic Q-factor | Full Q-factor | Full power amplification (dB) |
|---|---|---|---|---|---|---|---|---|
| (a) | $\infty$ | $10^{-9}$ | 0.077 | 2.57 | 24.6 | $5 \cdot 10^8$ | $3.82 \cdot 10^8$ | $-0.004$ |
| (b) | $1.05 v_0$ | $10^{-9}$ | 0.077 | 4.04 | 38.6 | $5 \cdot 10^8$ | $3.82 \cdot 10^8$ | $-0.004$ |
| (c) | $0.01 v_0$ | $10^{-9}$ | 0.077 | $-0.026$ | $-0.25$ | $5 \cdot 10^8$ | $3.82 \cdot 10^8$ | 0.19 |
| (d) | $0.001 v_0$ | $5 \cdot 10^{-9}$ | 0.39 | $-0.0026$ | $-0.025$ | $10^8$ | $9.4 \cdot 10^7$ | $-0.053$ |
| (e) | $0.001 v_0$ | $4 \cdot 10^{-9}$ | 0.31 | $-0.0026$ | $-0.025$ | $1.25 \cdot 10^8$ | $1.16 \cdot 10^8$ | 1.13 |
| (f) | $0.001 v_0$ | $3 \cdot 10^{-9}$ | 0.23 | $-0.0026$ | $-0.025$ | $1.67 \cdot 10^8$ | $1.51 \cdot 10^8$ | 24.3 |

**Table 1**. Parameters of plots shown in Fig. 9.

We assume that the input light frequency corresponds to the exact optical resonance, $\omega_0 T = 2\pi N_0$, and show that the offset $\Delta\omega_p$ can be set to arrive at the required transmission bandwidth $\Delta\omega_B$ of the optical resonator. For small offsets considered, $\Delta\omega_p T \ll 1$, we can replace the sum in Eq. (51) by an integral and calculate it by the stationary phase method. Calculations detailed in Appendix E, show that the transmission band of the resonator can be localized within the frequency band

$$-\frac{\Delta\omega_B}{2} + \omega_0 \leq \omega \leq \frac{\Delta\omega_B}{2} + \omega_0 \tag{55}$$

with the bandwidth

$$\Delta\omega_B = \frac{2|\Omega_p|\omega_p}{|\Delta\omega_p|T}. \tag{56}$$

In our numerical modelling, we set the modulation frequency equal to $\omega_p = 30$ GHz and modulation index to $\Delta n_p = 2 \cdot 10^{-4} n_0$ where $n_0 = 2.2$ is the refractive index of the lithium niobate. The modulation frequency offset $\Delta\omega_p$ (Eq. 54)) is chosen to arrive at the resonator bandwidth close to $\Delta\omega_B = 10$ THz (Eq. (56)), which includes up to 1000 comb lines. Figs. 8(a)-(f) show the results of our calculations for a lithium niobate racetrack resonator with the half-waveguide length close to $L = 2.27$ mm, which is assumed to be equal to the modulation length. The microresonator waveguide length $2L = 4.54$ mm was chosen to arrive at the condition close to the modulation resonance $\omega_p T = 2\omega_p L/v_0 = 2\pi$. The system parameters common for all plots are shown on the top of Fig. 9, while the parameters specific for each of the plots are shown on the top of individual plots (a)-(f) and more completely summarized in Table 1.

To estimate a feasible *maximum* amplification effect in the case of instantaneous ($v_p = \infty$) close to synchronous ($v_p = 1.05v_0$), and marginally small phase velocity ($v_p = 0.01v_0$) modulations, we set the imaginary part of refractive index in Figs. 8(a), (b), and (c) equal to $\eta_0 = 10^{-9} n_0$ corresponding to the material loss 0.077 dB/m (Table 1). In particular, in Fig. 9(a) we consider the instantaneous modulation with $v_p = \infty$. In this case, the modulation index found from Eq. (21) is $\Omega_p = 3.86$ and the modulation frequency offset corresponding to the bandwidth $\Delta\omega_B = 10$ THz found from Eq. (43) is $\Delta v_p = \Delta\omega_p/2\pi = 24.6$ MHz. The spectrum shown in Fig. 9(b) corresponds to $v_p = 1.05v_0$, i.e., is close to the synchronous modulation. In both cases, the effect of amplification appeared to be negligible. Next, in Fig. 10(c), we consider the case of small traveling wave phase velocity, $v_p = 0.01v_0$, corresponding to much smaller $\Omega_p = -0.039$ and $\Delta v_p = -0.25$ MHz. In this case, the effect of the amplification factor $F_m$ defined by Eq. (27) entering Eq. (51) is clearly visible. This factor leads to the attenuation of the transmission spectrum for negative $m$ and its amplification for positive $m$. However, the total amplification of the input wave power is still small, $P_{av}^{(out)} = 0.19$ dB. The bandwidth of transmission spectra shown in Figs. 8(a), (b), and (c) accurately coincide with the value predicted by Eq. (43).

The behavior of transmission spectrum shown in Fig. 9(c) suggests that a greater amplification can be achieved with a smaller traveling wave velocity $v_p$. The introduced amplification should overcome the material loss $\eta_0$, which was set to a very small value in Fig. 9(a), (b), and (c). To illustrate the feasibility of amplification, in Figs. 8(d), (e), and (f), we set $v_p = 0.001v_0$. This value still satisfies the eikonal condition of slowness of the modulation in space, $\frac{\omega_p}{v_p} \ll \frac{\omega_0}{v_0}$ (see Eq. (4)). However, the value of $v_p$ cannot be reduced further within the correctness of the eikonal approximation. In Section IV.B, we will overcome this limitation using the perturbation theory approach developed in Section III and show that the effect of modulation with slow phase velocity $v_p$ can be dramatically enhanced at $\frac{\omega_p}{v_p} \cong 2\frac{\omega_0}{v_0}$.

Figs. 8(d), (e), and (f) show that, to arrive at substantial amplification of light, the transmission loss of the waveguide of the optical resonator should be exceptionally small. In these figures, the waveguide loss $\eta_0$ is set to $5 \cdot 10^{-9} n_0$ (Fig. 9(d)), $4 \cdot 10^{-9} n_0$ (Fig. 9(e)), and $3 \cdot 10^{-9} n_0$ (Fig. 9(f)). These values correspond, respectively, to the light attenuation of 0.39, 0.31, and 0.23 dB/m and intrinsic resonator Q-factors of $10^8$, $2.5 \cdot 10^8$, and $1.67 \cdot 10^8$ (Table 1). The latter values are comparable to the attenuation of 0.2 dB/m and intrinsic Q-factor of $1.6 \cdot 10^8$ recently achieved in a lithium niobate waveguide [54]. Fig. 9(d) shows that the modulation-induced amplification is not sufficient to overcome the attenuation of 0.39 dB/m. However, Fig. 9(e) demonstrates a small full amplification of 1.13 dB for the waveguide loss of 0.31 dB/m. Finally, for a waveguide with propagation loss of 0.23 dB/m achieved in [54], Fig. 9(f) demonstrates a substantial amplification of the input wave power by 24.3 dB as well as the amplification of individual comb line powers.

We notice the characteristic reduction of the width of transmission band and its blue-shifted deformation for the cases demonstrating significant amplification (Fig. 9(d), (e), and (f) compared to the cases with a larger traveling wave velocity (Fig. 9(a), (b), and (c)). This effect is caused by tilting the transmission spectrum imposed by the amplification factor $F_m$ (Eq. (27)) whose logarithm is a linear function of the comb number, $\ln(F_m) \sim m$.

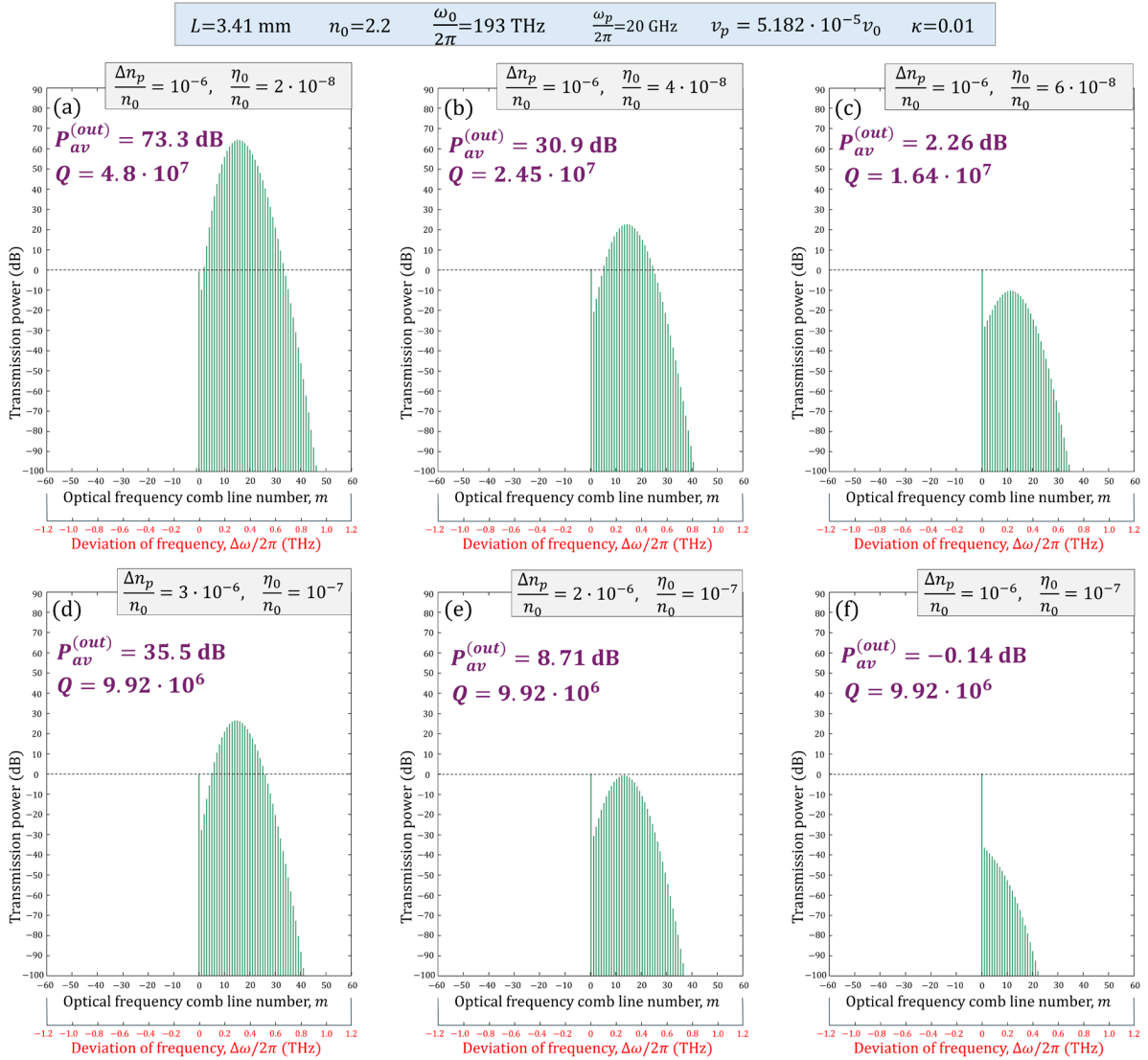

**Fig. 10.** The resonant transmission power spectra for a racetrack resonator modulated by a traveling wave which is applied and does not attenuate along the resonator length $L$. The system parameters are indicated on the top of the figure. Plots (a)-(f) correspond to different refractive index modulation amplitude $\Delta n_p$ and waveguide propagation losses $\eta_0$. These and other parameters of the plots are detailed in Table 2.

## B. Gigantic narrowband resonant amplification near the phase matching condition

We found in the previous Section IV.A that light amplification can be achieved within a relatively small bandwidth $\Delta \omega_B \ll \omega_0$ provided that the phase velocity of the modulating wave $v_p$ is small enough. However, the eikonal approximation used did not allow us to consider sufficiently small values of $v_p$ since, according to Eq. (4), the condition of slowness of modulation in space restricts these values to

$v_p \gg \omega_p v_0/\omega_0$. Remarkably, the latter restriction can be withdrawn for the parameters of our interest. Indeed, as shown in Section IV.A, the values of modulation index $\Omega_p(L)$ required to arrive at the substantial amplification of light are small, $|\Omega_p(L)| \ll 1$. Then, solution of the wave equation, Eq. (1), can be found by the regular perturbation theory developed in Section III rather than by the eikonal approximation.

Using the solution of the wave equation determined in Section III for modulation with attenuation (Fig. 8(b)), we determine the output amplitude $E_{out}(t)$ following Eqs. (46)-(48) where the expression for function $A(t)$ is now modified to

$$A(t) = \exp\left[i\omega_0 T - \frac{\eta_0}{n_0}\omega_0 T + i\tilde{\Omega}_p(L)\cos\left(\omega_p t + i\Theta(L)\right)\right]. \tag{57}$$

Here functions $\tilde{\Omega}_p(x)$ and $\Theta(x)$ are defined by Eqs. (37)-(39). It is now straightforward to determine the output transmission amplitude $E_{out}(t)$ by comparing the expressions for $A(t)$ in Eqs. (49) and (57). It follows from this comparison that the comb spectral amplitudes $U_m^{(c)}$ of $E_{out}(t)$ can be determined from Eq. (51) after the substitutions:

$$\Omega_p(L) \to \tilde{\Omega}_p(L), \quad iG_p \to i\tilde{G}_p(L). \tag{58}$$

| Resonator half-length $L$ (mm) | Refractive index $n_0$ | Light frequency $\frac{\omega_0}{2\pi}$ (GHz) | Travelling wave phase velocity $v_p$ (m/s) | Modulation frequency $\frac{\omega_p}{2\pi}$ (GHz) | Coupling coefficient $\kappa$ |
|---|---|---|---|---|---|
| 3.41 | 2.2 | 193,000 | 7068 | 20 | 0.01 |

| Plot | Refractive index modulation $\Delta n_p$ | Relative material loss $\eta_0/n_0$ | Material loss $20\log(e) \times \eta_0 \omega_0/c$ (dB/m) | Modulation index $\Omega_p$ | Deviation from modulation resonance $\frac{\Delta \omega_p}{2\pi}$ (MHz) | Intrinsic Q-factor | Full Q-factor | Full power amplification (dB) |
|---|---|---|---|---|---|---|---|---|
| (a) | $2.2 \cdot 10^{-6}$ | $2 \cdot 10^{-8}$ | 1.54 | $-6.4 \cdot 10^{-5}$ | $-1.02$ | $5 \cdot 10^7$ | $4.8 \cdot 10^7$ | 73.3 |
| (b) | $2.2 \cdot 10^{-6}$ | $4 \cdot 10^{-8}$ | 3.09 | $-6.4 \cdot 10^{-5}$ | $-1.02$ | $2.5 \cdot 10^7$ | $2.45 \cdot 10^7$ | 30.9 |
| (c) | $2.2 \cdot 10^{-6}$ | $6 \cdot 10^{-8}$ | 4.63 | $-6.4 \cdot 10^{-5}$ | $-1.02$ | $1.67 \cdot 10^7$ | $1.64 \cdot 10^7$ | 2.26 |
| (d) | $6.6 \cdot 10^{-6}$ | $10^{-7}$ | 7.72 | $-1.92 \cdot 10^{-4}$ | $-3.06$ | $10^7$ | $9.92 \cdot 10^6$ | 35.5 |
| (e) | $4.4 \cdot 10^{-6}$ | $10^{-7}$ | 7.72 | $-1.28 \cdot 10^{-4}$ | $-2.04$ | $10^7$ | $9.92 \cdot 10^6$ | 8.71 |
| (f) | $2.2 \cdot 10^{-6}$ | $10^{-7}$ | 7.72 | $-6.4 \cdot 10^{-5}$ | $-1.02$ | $10^7$ | $9.92 \cdot 10^6$ | $-0.14$ |

**Table 2**. Parameters of plots shown in Fig. 10.

### 1. Modulation with zero attenuation

First, we consider the modulation with negligible attenuation, $\alpha = 0$ (Fig. 8(a)). The results of our calculations based on Eq. (51) corrected by the substitutions of Eq. (58) are shown in Fig. 10. The parameters of plots in this figure are summarized in Table 2. Fig. 10 demonstrates that, near the resonant value of $v_p$ chosen close to that defined by Eq. (41) and satisfying the condition of Eq. (40), the substantial amplification effect can be achieved for a much smaller modulation of refractive index and greater losses compared to those required away from this singularity and shown Fig. 9. Fig. 10(a) demonstrates the dramatic total amplification of $P_{av}^{(out)} = 73.3$ dB for a racetrack resonator with waveguide loss 1.54 dB/m, which is modulated with frequency $\omega_p = 2\pi \cdot 20$ GHz and modulation amplitude of only $\Delta n_p = 2.2 \cdot 10^{-6}$. In this case, the individual comb lines are significantly amplified as well. Fig. 10(b) shows that a waveguide with loss increased by a factor of two to 3.09 dB/m can still manifest large amplification with total power $P_{av}^{(out)} = 30.9$ dB and amplification of a fraction of individual comb lines. Further increase of the waveguide loss to 4.63 dB/m considered in Fig. 10(c) still results in total amplification of $P_{av}^{(out)} = 2.26$ dB though the absence of individual comb line amplification. Fig. 10(d) shows that the increase of refractive index modulation amplitude by a factor of three compared to that considered in Fig. 10(a), (b), and (c) to $\Delta n_p = 6.6 \cdot 10^{-6}$ (still dramatically small) and increase of the waveguide loss to 7.72 dB/m, leads to the total amplification of $P_{av}^{(out)} = 35.5$ dB as well as the individual comb line amplification. For the same loss, the amplification is reduced to $P_{av}^{(out)} = 8.71$ dB for smaller modulation amplitude $\Delta n_p = 4.4 \cdot 10^{-6}$ and completely vanishes for $\Delta n_p = 2.2 \cdot 10^{-6}$.

### 2. *Spatially attenuating modulation*

Next, in our numerical example of modulation induced amplification in an optical resonator considered in Fig. 11, we set the traveling wave velocity equal to the characteristic surface acoustic wave (SAW) velocity in lithium niobate, $v_p = 3500$ m/s, the modulation frequency equal to 9.9041 GHz (i.e., approximately satisfying Eq. (41)) and the modulation amplitude $\Delta n_p = 2 \cdot 10^{-5}$. For these $\omega_p$, $v_p$, and $\Delta n_p$, and the modulation attenuation $\alpha$ chosen in Figs 10 (a), (b), and (c), the condition of validity of our approximation, Eq. (40), is well satisfied. The half-length of the resonator satisfying the resonant modulation condition is now $L = 6.89$ mm. In the transmission spectrum plots shown in Fig. 11 we choose $\eta_0 = 6.5 \cdot 10^{-8} n_0$ corresponding to the material loss of 5 dB/m. The detailed values of all parameters are given in Table 3. In all plots of Fig. (10) the values $\alpha$ of modulation attenuation ensure that it is confined within the half resonator waveguide length $L$. It is seen from these plots that, for the attenuation values $\alpha = 500$ m$^{-1}$, 1000 m$^{-1}$, and 1500 m$^{-1}$ corresponding respectively to 21.7 dB/cm, 43.4 dB/cm, and 65.1 dB/cm, the light power can be amplified with the total amplification of 61 dB, 20.9 dB, and 1.71 dB.

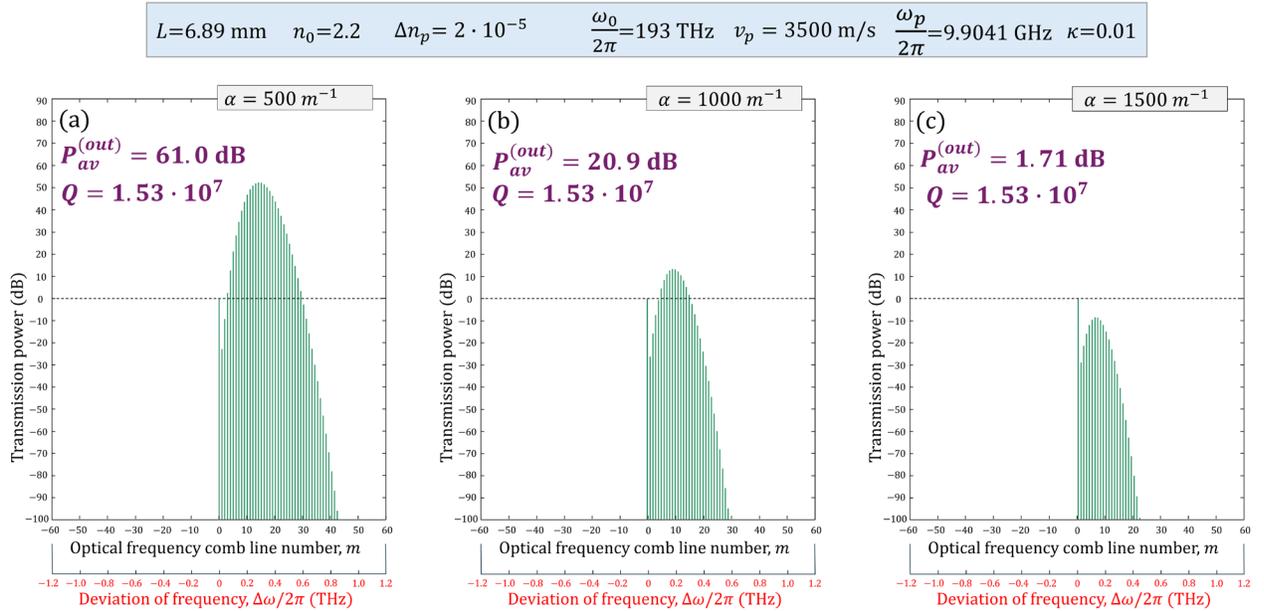

Fig. 11. The resonant transmission power spectra for a racetrack resonator modulated by attenuating traveling wave. The system parameters are indicated on the top of the figure. These and other parameters of the plots are detailed in Table 3.

| Resonator half-length $L$ (mm) | Refractive index $n_0$ | Light frequency $\frac{\omega_0}{2\pi}$ (GHz) | Travelling wave phase velocity $v_p$ (m/s) | Modulation frequency $\frac{\omega_p}{2\pi}$ (GHz) | Coupling coefficient $\kappa$ |
|---|---|---|---|---|---|
| 6.89 | 2.2 | 193,000 | 3500 | 9.9041 | 0.01 |

| Plot | Refractive index modulation $\Delta n_p$ | Relative material loss $\eta_0/n_0$ | Material loss $20\log(e) \times \eta_0\omega_0/c$ (dB/m) | Modulation index $\widetilde{\Omega}_p$ | Deviation from modulation resonance $\frac{\Delta\omega_p}{2\pi}$ (MHz) | Intrinsic Q-factor | Full Q-factor | Attenuation coefficient of modulation $\alpha$ ($m^{-1}$) | Attenuation coefficient of modulation $\alpha$ (dB/cm) | Full power amplification (dB) |
|---|---|---|---|---|---|---|---|---|---|---|
| (a) | | | | | | | | 500 | 21.7 | 61.0 |
| (b) | $2 \cdot 10^{-5}$ | $6.5 \cdot 10^{-8}$ | 5.02 | $(2.8 - 1.9i) \cdot 10^{-4}$ | 5.41 | $1.54 \cdot 10^7$ | $1.53 \cdot 10^7$ | 1000 | 43.4 | 20.9 |
| (c) | | | | | | | | 1500 | 65.1 | 1.71 |

Table 3. Parameters of plots shown in Fig. 11.

## V. EXPERIMENTAL FEASIBILITY OF AMPLIFICATION

Current progress in the research and development of lithium niobate optical microresonators with exceptionally small losses [53, 54, 63] and eigenfrequency dispersion [64, 65], as well as in the design of microscopic RF electromagnetic and acoustic traveling wave generators [66-70], suggests that the system parameters required for the substantial amplification of light by electromagnetic and acoustic waves with dramatically smaller frequencies are feasible. In this Section, we compare the microresonator and modulation parameters considered in Section IV with those experimentally achievable.

*Waveguide propagation loss.* The condition of amplification demonstrated in Section IV imposes a significant upper bound on the waveguide propagation loss of an optical resonator. This restriction can be relaxed by decreasing the ratio of phase velocities $v_p/v_0$ and increasing the modulation amplitude $\Delta n_p$. The phase velocity considered in Section IV.A is $v_p = 0.001v_0 \cong 136300$ m/s, while the modulation frequency and amplitude are $\omega_p = 2\pi \cdot 30$ GHz and $\Delta n_p = 4.4 \cdot 10^{-4}$. Practically, it is challenging to achieve so large modulation frequency and amplitude simultaneously. Here, these values were chosen to arrive at the smallest practically achievable waveguide loss of 0.23 dB/m required for the substantial amplification of light (see Fig. 9(f) and Table 1). A lithium niobate resonator with dramatically small waveguide loss 0.34 dB/m approaching the bulk material loss was demonstrated recently by chemo-mechanical waveguide polishing in Ref. [53]. The waveguide loss as small as 0.2 dB/m was demonstrated in Ref. [54] by post-fabrication annealing in oxygen atmosphere. Remarkably, the waveguides, which support the substantial amplification of light considered in Section IV.B for much smaller $v_p = 7068$ m/s and $v_p = 3500$ m/s, possess much easier to realize loss in the range 1-7 dB/m and much smaller modulation amplitude $\Delta n_p$ in the range $10^{-6}$ - $10^{-5}$ (see Figs. 9(a), (b), (d), (e), Figs 10 (a), (b), (c) and Tables 2, 3. Indeed, the authors of Ref. [63] developed a robust method for fabrication of the thin-film lithium niobate waveguides based on the dry reactive ion etching with an ultra-low propagation loss which can be as small as 1.3 dB/m. Thus, the waveguides with losses required to experimentally realize the amplification with parameters of the transmission spectrum shown in Figs. 10 and 11 has been experimentally demonstrated.

*Resonator eigenfrequency dispersion.* The waveguide dispersion can be optimized to arrive at the smallest possible eigenfrequency dispersion. In contrast to the optimization for the optical frequency comb spectrum generated by optical microresonators commonly targeted at the largest possible bandwidth [30, 64], here we are interested in the accurate minimization of dispersion along a finite bandwidth $\Delta\omega_B$. In Section IV.A, we have $\Delta\omega_B = 10$ THz (Fig. 9(f)), in Section IV.B.1, we have $\Delta\omega_B = 0.8$ THz (Fig. 10(a), (b), (d), and (e)), and, in Section IV.B.2 we have $\Delta\omega_B = 0.4$ THz (Fig. 11(a), (b), and (c)). The eigenfrequency dispersion of microresonators is characterized by the deviation from linear dependence $\delta\omega(\Delta\omega) = \omega_m - \omega_0 - m\omega_p$ of their spectral series $\omega_m$. Here $\Delta\omega$ is the continuous extrapolation of $m\omega_p$ and $m = \text{int}(\Delta\omega/\omega_p)$ (see, e.g., [65, 71]). For the light frequency $\omega_0$ in the vicinity of the $\delta\omega(\Delta\omega)$ minimum, the amplification power will approach the values determined above in Sections IV.A and B if $\delta\omega(\Delta\omega_B)$ is much smaller than the resonance width $\Delta\omega_{res} = \omega_0/Q$ where the quality factor $Q$ is determined by Eq. (53), $|\delta\omega(\Delta\omega_B)| \ll \Delta\omega_{res}$. For the microresonator and modulation parameters leading to the amplification with the transmission spectrum of Fig. 9(f), we have $Q \sim 10^8$ (Table 1) and resonance width $\Delta\omega_{res} \sim 2$ MHz. To our knowledge, such a small microresonator dispersion has not been demonstrated to date. However, for the modulation and microresonator parameters considered in Section IV.B (Tables 2, 3), the resonance width is $\Delta\omega_{res} \sim 20$ MHz. Remarkably, the value of deviation $\delta\omega(\Delta\omega_B)$ achieved in Ref. [65] for lithium niobate and in Ref. [71] for silicon nitride microresonators is smaller than 10 MHz over the bandwidth $\Delta\omega_B = 0.8$ THz (Fig. 10) and much smaller than 10 MHz over the bandwidth $\Delta\omega_B = 0.4$ THz (Fig. 11). Thus, the microresonators with eigenfrequency dispersion required to experimentally realize the amplification with parameters of the transmission spectrum shown in Fig. 11 has been experimentally demonstrated.

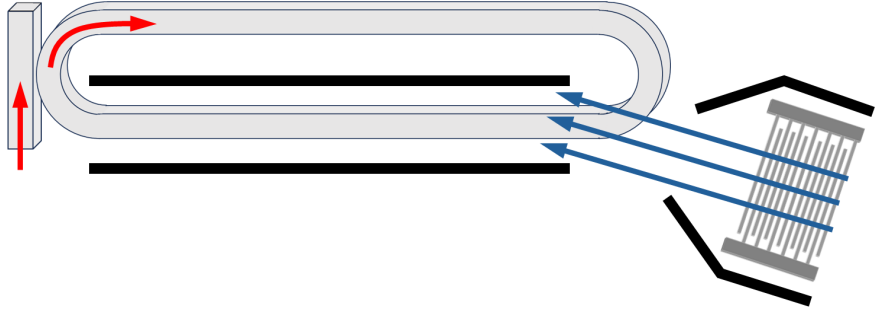

**Fig. 12.** Illustration of a racetrack resonator modulated by a SAW.

*Modulation methods.* Different approaches have been developed for the effective optical waveguide RF modulation (see [25, 26, 33, 60, 66-70] and references therein). The simplest design is represented by a spatially uniform capacitor, which can generate instantaneous modulation $\Delta n_p \cos(\omega_p t)$ corresponding to $v_p = \infty$ (see e.g., [33, 64]). A traveling wave refractive index modulation $\Delta n_0 \cos\left(\omega_p(t - x/v_p)\right)$ can also be introduced by an RF wave propagating parallel to the optical waveguide (see, e.g., [66, 25-27]). This approach is beneficial for the modulation of photonic circuits with a traveling wave having the phase velocity $v_p$ comparable or equal to the phase velocity of light $v_0$. For applications of our concern, we are interested in relatively small traveling wave phase velocities $v_p \ll v_0$. For this purpose, the surface acoustic waves (SAWs) and bulk acoustic waves generated by an interdigital transducer (IDT) can be used [36, 67-70]. SAWs modulate the refractive index of an optical waveguide through the elasto-optics effect. An IDT tilted with respect to an optical waveguide by angle $\theta$ generates SAW propagating along the waveguide with the phase velocity $v_p = v_{sound}/\sin(\theta)$, where $v_{sound}$ is the speed of sound in the material. The SAWs can be enhanced and aligned along an optical waveguide by the combination of a focused IDT and an acoustic waveguide as illustrated in Fig. 12 (see, e.g., [68, 69]). The characteristic refractive index variation induced by SAWs generated by an IDT in lithium niobate is estimated as $\Delta n_p \sim \frac{1}{2} n_0^3 (r_{33} - p_{33} d_{33}) V/w$, where $V$ is the voltage applied to the IDT, $w$ is the characteristic IDT finger separation, and we set the electro-optic coefficient $r_{33} = 30$ pm/V, the photoelastic coefficient $p_{33} = 0.1$, the piezoelectric coefficient $d_{33} = 6$ pm/V. Assuming $\frac{V}{w} \sim 1$ V/μm we find $\Delta n_p \sim 10^{-4}$ at the IDT position. SAWs experience strong exponential attenuation with distance from the IDT dramatically increasing with frequency. In lithium niobate, at modulation frequency $\omega_p \cong 10$ GHz the SAW attenuation has the smallest experimentally achieved value of $\cong 20$ dB/cm [59] corresponding to the attenuation $\alpha \cong 460$ m$^{-1}$ m in our model described by Eq. (33). Consequently, the attenuations of $\alpha = 500$ m$^{-1}$, $1000$ m$^{-1}$ and $1500$ m$^{-1}$ and initial refractive index modulation amplitude of $\Delta n_p = 2 \cdot 10^{-5}$ in the transmission spectra shown in Fig. 11 are experimentally feasible.

In a more general case, an advanced design of the modulators based on elasto-optic and electro-optic effects is required to arrive at the phase velocity, frequency, and spatial distribution of modulation leading to the amplification of light. Commonly, a single tilted IDT generating a traveling wave with velocity $v_p \ll v_0$ is insufficient to generate the amplification of light due to the rapid SAW attenuation in space. However, a combination of in-phase tilted IDTs distributed along the optical waveguide and generating properly aligned SAWs may solve the problem. Alternatively, modulation of refractive index of an optical waveguide through a combination of bulk acoustic wave and Pockels electro-optic effects can be introduced by the RF waveguides spatially modulated with the period $d$ and aligned along the optical waveguide. In this design, one RF waveguide introduces the optical waveguide's refractive index modulation equal to $\Delta n_0 \cos\left(\omega_p(t - x/v_{p0})\right) \cos(2\pi x/d)$. In turn, another RF waveguide introduced modulation $\Delta n_0 \sin\left(\omega_p(t - x/v_{p0})\right) \sin(2\pi x/d)$, which is phase shifted in time from the first one by $\pi/2$. The superposition of these modulations yields the traveling wave $\Delta n_0 \cos(\omega_p(x/v_p - t))$ with the phase velocity $v_p = \left(\frac{2\pi}{d\omega_p} + \frac{1}{v_{p0}}\right)^{-1}$ which can be small for a small IDT period $d$. For example, for the case considered in Section IV.A and $v_{p0} \sim v_0$, we have $d \cong 2\pi v_p/\omega_p = 4.55$ µm. Realization of such complex modulation structures is challenging since their proximity to the optical waveguide leading to the enhancement of modulation strength should be compromised with their effect on the optical waveguide loss.

## VI. DISCUSSION

We investigated the propagation of light with frequency $\omega_0$ through optical waveguides and racetrack resonators modulated by a traveling wave with relatively small frequency $\omega_p \ll \omega_0$ using the eikonal (semiclassical, WKB) approximation [6, 13, 14] (Sections II and IV.A) and a regular perturbation theory (Sections III and IV.B). In Section V, we compared the parameters of waveguides and resonators used in our numerical modeling with those feasible experimentally and found that a dramatically large amplification of light is possible to achieve in a lithium niobate racetrack resonator modulated by an acoustic traveling wave.

Section II of the paper was dedicated to the analysis of propagation of light in ideally dispersionless waveguides under the modulation of a traveling wave with a spatially uniform amplitude. We derived a general eikonal expression for the transmission amplitude similar to that derived several decades ago in application to the propagation of electromagnetic waves in transmission lines [5, 6]. Consequently, most of the results presented in this Section either resemble or complement the previously known results being now applied to the propagation of light. In particular, we showed that the effect of the traveling wave modulation can be significantly enhanced if modulation takes place along a sufficiently large waveguide length $L$ and in a close vicinity of the completely synchronous condition $v_p = v$. The determined transmission amplitude is periodic in time. However, its dependence on modulation length $L$ becomes aperiodic and growing for sufficiently small material losses if $|\mu| > 1$ and remains quasiperiodic if $|\mu| < 1$ [5, 6, 8, 10]. We also showed that synchronous modulation is not advantageous compared to the commonly used instantaneous modulation for a relatively small modulation length $L$ of several tens of microns (see Figs. 5(a) and (b) and Figs. 6(c), (d) and (f)).

One of the major goals of this paper was understanding the feasibility of light amplification by a low frequency modulation of realistic optical waveguides and resonators. Having in mind realistic applications, we focused on the situations when light is propagating within a relatively small bandwidth $\Delta \omega_B(L) \ll \omega_0$. We found that, for a relatively large phase velocity of modulating wave, $v_p \gg v_0$, or for comparable velocities, $v_p \sim v_0$, the averaged over time amplification $P_{av}(L)$ is always small being proportional to

$(\Delta\omega_B(L)/\omega_0)^2$ (see Eq. (31)). The situation becomes different for $v_p \ll v_0$ when Eq. (31) suggests that a moderate amplification of light may be possible within a small bandwidth $\Delta\omega_B(L)$. In Section III we considered the latter case and found that the amplification of light can become noticeable near the Brillouin phase marching condition $v_0\omega_p = 2v_p\omega_0$ (see Eqs. (42) and (45)). However, we also found that such an amplification is currently very challenging to achieve experimentally due to the large attenuation and small power of the surface acoustic waves possessing the required small phase velocities $v_p$.

In contrast, we showed in Section IV that dramatic amplification of light is possible in a modulated optical racetrack resonator. Generally, a larger amplification can be achieved with a larger modulation frequency $\omega_p$ and amplitude $\Delta n_p$ and a smaller resonator waveguide loss $\eta_0$. To demonstrate the feasibility of amplification, we noticed that the transmission bandwidth $\Delta\omega_B$ of a resonator can be controlled and made small by increasing the offset of frequency $\omega_p$ from the exact modulation resonance. We found that, in the eikonal approximation, the amplification of individual comb lines increases with the parameter $mv_0\omega_p/v_p\omega_0$, where $m$ is the frequency comp line number (see the expression for $G_p$ in Eqs. (51) and (27)). For the small modulation frequencies of our interest, $\omega_p \ll \omega_0$, this parameter increases with a decrease of the phase velocity of the modulating wave $v_p$. Next, we showed that, similar to the propagation in optical waveguides considered in Section III, the amplification effect can be dramatically enhanced near the Brillouin phase marching condition $v_0\omega_p = 2v_p\omega_0$. As an example, we demonstrated that the amplification of the full power of light as much as 60 dB can be achieved within the bandwidth $\Delta\omega_B = 400$ GHz in a racetrack resonator with the waveguide half-length 6.89 mm and loss 5 dB/m modulated by acoustic waves with velocity $v_p = 3500$ m/s, initial modulation index $\Delta n_p = 2 \cdot 10^{-5}$, and attenuation 21.7 dB/cm (Fig. 11(a) and Table 3). We showed in Section V that all the required resonator waveguide and modulation parameters are currently feasible. We suggest that further optimization of the parameters considered in our modeling, combined with advanced design of the modulation components and microresonator configuration, could enable the practical realization of the proposed acoustic light amplifier, achieving the outstanding performance predicted here.

**Acknowledgement:** This research was supported by the Engineering and Physical Sciences Research Council (Grant EP/W002868/1), Leverhulme Trust (Grant RPG-2022-014), and Royal Society (Grant IES\R1\231250). The author is grateful to A. A. Fotiadi and S. K. Turitsyn for fruitful and inspiring discussions.

**Appendix A.**

**Eikonal (WKB) approximation**

We assume that the refractive index is a slow function of time and coordinates and formally introduce slow coordinate and time, $\zeta = \varepsilon x$ and $\tau = \varepsilon t$, where $\varepsilon \sim \omega_p/\omega_0 \ll 1$. We look for the solution of Eq. (1) in the form [13, 14]:

$$E(x,t) = \left(U_0(x,t) + \varepsilon U_1(x,t) + \ldots\right) \exp\left(\frac{i}{\varepsilon} S(x,t)\right). \tag{A1}$$

Substituting Eq. (A1) into Eq. (1) and expanding the result in powers of $\varepsilon$, we arrive at a series of coupled equations for $U_m(x,t)$ and $S(x,t)$. In the zero order in $\varepsilon$, we obtain the equation for the eikonal $S(x,t)$ which determines the phase of solution:

$$n^2(x,t)S_t^2 - c^2 S_x^2 = 0. \tag{A2}$$

This equation is reduced to the linear equation

$$n(x,t)S_t + cS_x = 0, \tag{A3}$$

where it is assumed that the speed of light $c$ can have positive or negative sign. Once the solution $S(x,t)$ of Eq. (A3) is found, the amplitude terms $U_m(x,t)$ are determined from linear equations that can be solved successively. In particular, the zero-order term $U_0(x,t)$ of the amplitude of solution is found from the equation:

$$n(x,t)U_{0t} + cU_{0x} + \left(\frac{3}{2}n_t + \frac{c}{2n}n_x\right)U = 0. \tag{A4}$$

For the case of the traveling wave refractive index defined by Eq. (2), solution of eikonal Eq. (A2) for the field phase and Eq. (A4) for the field amplitude can be found by the introduction of variables

$$\begin{aligned} t' &= t - \frac{x}{v_0}, \\ x' &= x, \end{aligned} \tag{A5}$$

where $v = c/n_0$ is the phase velocity of light (Fig. 1). Then, the refractive index in Eq. (2) depends on $t'$ only and Eqs. (A4) and (A5) can be rewritten down as

$$(a + b\cos(\omega_p t'))S_{t'} + vS_{x'} = 0, \tag{A6}$$

$$(a + b\cos(\omega_p t'))U_{0t'} + vU_{0x'} - \delta(t')U = 0, \tag{A7}$$

where

$$a = 1 - \frac{v_0}{v_p} + i\frac{\eta_0}{n_0}, \quad b = \frac{\Delta n_0}{n_0},$$

$$\delta(t) = b\omega_p \sin(\omega_p t)\frac{2 + b + 3b\cos(\omega_p t)}{2(1 + b\cos(\omega_p t))}. \tag{A8}$$

The general solutions of Eqs. (A3) and (A4) expressed through the original variables $x$ and $t$ are [72]

$$S(x,t) = \Phi_0(\xi(x,t)), \tag{A9}$$

$$U_0(x,t) = \Phi_1(\xi(x,t))W\left(t - \frac{x}{v_p}\right),$$

$$W(t) = \exp\left(\int_0^t \frac{\delta(t)dt}{a + b\cos(\omega_p t)}\right) = \frac{(a+b)\sqrt{1+b}}{(a + b\cos(\omega_p t))\sqrt{1 + b\cos(\omega_p t)}}. \tag{A10}$$

Here $\Phi_k(\xi)$ are arbitrary functions determined by the boundary and initial conditions and

$$\xi(x,t) = x - v \int_0^{t-\frac{x}{v_p}} \frac{dt}{a + b\cos(\omega_p t)}$$

$$= x - \frac{2v}{\omega_p \sqrt{a^2 - b^2}} \Xi\left(\sqrt{\frac{a-b}{a+b}}, \frac{\omega_p}{2}\left(t - \frac{x}{v_p}\right)\right). \tag{A11}$$

Here, function $\Xi(x, y)$ is the smooth continuation of $\arctan(x \cdot \tan(y))$ as a function of $y$ which is convenient to calculate as

$$\Xi(x,y) = \int_0^y \frac{\partial}{\partial y}\left(\arctan(x, \tan(y))\right) dy = \int_0^y \frac{dy}{x^2 \sin^2(y) + \cos^2(y)} \tag{A12}$$

Ignoring the reflected wave, we determine the asymptotic solution of Eq. (1) corresponding to the boundary condition at $x = 0$ (Fig. 1)

$$E^{(in)}(x,t) = \exp\left[i\omega\left(\frac{x}{v_0} - t\right)\right], \quad x < 0, \tag{A13}$$

separating it into the boundary conditions for $S(x, t)$ and $U_0(x, t)$:

$$S(0,t) = -\omega t, \tag{A14}$$

$$U_0(0,t) = 1. \tag{A15}$$

Following the approach of Ref. [5], we introduce function $\bar{t}(\bar{\xi})$ inverse to function $\bar{\xi}(t) = \xi(t, 0)$ which is found from Eq. (A11) as

$$\bar{t}(\bar{\xi}) = -\frac{2}{\omega_p} \Xi\left(\sqrt{\frac{a+b}{a-b}}, \frac{\omega_p}{2v_0}\sqrt{a^2 - b^2}\bar{\xi}\right) \tag{A16}$$

where, again, function $\Xi(x, y)$ is the smooth continuation of $\arctan(x \cdot \tan(y))$ as a function of $y$ defined by Eq. (A12). Using Eqs. (A11)-(A16), we find:

$$S(x,t) = -\omega \bar{t}(\xi(x,t)) = \frac{2\omega}{\omega_p} \Xi\left(\sqrt{\frac{a+b}{a-b}}, \frac{\omega_p}{2v_0}\sqrt{a^2 - b^2}\bar{\xi}\right) \tag{A17}$$

and

$$U_0(x,t) = \frac{W\left(t - \frac{x}{v_p}\right)}{W(\bar{t}(\xi(x,t)))} = \frac{\left(a + b\cos(\omega_p \bar{t}(\xi(x,t)))\right)\sqrt{1 + b\cos(\omega_p \bar{t}(\xi(x,t)))}}{\left(a + b\cos\left(\omega_p\left(t - \frac{x}{v_p}\right)\right)\right)\sqrt{1 + b\cos\left(\omega_p\left(t - \frac{x}{v_p}\right)\right)}}. \tag{A18}$$

## Appendix B.

## The outgoing wave for the completely synchronous and lossless case $v_p = v_0$ and $\eta_0 = 0$

In the synchronous lossless case, $v_p = v_0$ and $\eta_0 = 0$ (i.e., $a = 0$), Eqs. (A17) and (A18) are simplified. Setting $\Xi(x,y) = \arctan(x \cdot \tan(y))$, we find

$$S_0(x,t) = \frac{2\omega}{\omega_p} \arctan\left( i \tan\left( \frac{i\omega_p bx}{2v_0} - \arctan\left( i \tan\left( \frac{\omega_p}{2}\left(t - \frac{x}{v_p}\right)\right)\right)\right)\right)$$

(B1)

$$= \frac{2\omega}{\omega_p} \arctan\left( \frac{\tanh\left(\frac{b\omega_p x}{2v_0}\right) - \tan\left(\frac{\omega_p}{2}\left(t - \frac{x}{v_p}\right)\right)}{1 - \tanh\left(\frac{b\omega_p x}{2v_0}\right)\tan\left(\frac{\omega_p}{2}\left(t - \frac{x}{v_p}\right)\right)}\right).$$

Next, we simplify Eq. (A17) for the amplitude $U_0(x,t)$ under the same assumption $a = 0$ assuming $\Delta n_p/n_0 \ll 1$. As the result we have:

$$U_0(x,t) = \frac{\cos\left(\omega_p \bar{t}(\xi(x,t))\right)}{\cos\left(\omega_p\left(t - \frac{x}{v_p}\right)\right)}.$$

(B2)

where

$$\cos\left(\omega_p \bar{t}(\xi(x,t))\right) = \left(\cosh\left(\frac{b\omega_p}{v_0}\xi(x,t)\right)\right)^{-1}$$

(B3)

Then, from Eqs. (B2) and (B3),

$$U_0(x,t) = \frac{1}{\cosh\left(\frac{b\omega_p x}{v_0}\right) - \sin\left(\omega_p\left(t - \frac{x}{v_p}\right)\right)\sinh\left(\frac{b\omega_p x}{v}\right)}.$$

(B4)

Averaging the output power, $P(x,t) = U_0(x,t)^2$, over time we find:

$$P_{av}(x) = \frac{\omega_p}{2\pi}\int_0^{2\pi/\omega_p} U_0(x,t)^2 dt = \cosh\left(\frac{b\omega_p x}{v_0}\right).$$

(B6)

It is seen from Eq. (B4) that for a sufficiently large modulation length $x$ corresponding to $\exp\left(\frac{b\omega_p}{v_0}\right) \gg 1$ the amplitude $U_0(x,t)$ rapidly changes with time in small intervals $\sin\left(\omega_p(t - x/v_p)\right)$ is close to unity. We determine the position $t_{max}$ of the maximum slope of $U_0(x,t)$ as a function of time by finding the zeros of its second derivative. Then, the condition of validity of the eikonal approximation, $\frac{\partial}{\partial t}U_0(x,t)\big|_{t=t_{max}} \ll \omega_0$ yields Eq. (12) of the main text.

**Appendix C.**

**The spectral bandwidth for the completely synchronous and lossless case $v_p = v$ and $\eta_0 = 0$**

We determine the transmission bandwidth by calculating the integral for the frequency comb amplitude given by Eq. (15) using the stationary phase method. For briefness, we introduce notations:

$$t_{xt} = \tan\left(\omega_p\left(\frac{x}{v_p} - t\right)\right), \quad t_x = \tanh\left(\frac{b\omega_p x}{2v_0}\right). \tag{C1}$$

The stationary phase time is then found by zeroing the derivative of the phase in the exponent of the integral of Eq. (15) where $E(x,t)$ is determined by Eq. (11):

$$\frac{\omega_0\left(t_x^2 - 1\right)\left(t_{xt}^2 + 1\right)}{t_x^2 t_{xt}^2 + t_x^2 + 4t_x t_{xt} + t_{xt}^2 + 1} + \omega_0 - \omega_p n = 0. \tag{C2}$$

From here, we find

$$t_{xt}^{\pm} = \frac{\left(\varsigma_+ \varsigma_-\right)^{1/2}\left(t_x^2 - 1\right)^{1/2} \pm 2t_x\left(\frac{\omega_0}{\omega_p} - n\right)}{\left(n - 2\frac{\omega_0}{\omega_p}\right)t_x^2 - n}, \tag{C3}$$

$$\varsigma_{\pm} = n + \left(n \pm 2\frac{\omega_0}{\omega_p}\right)t_x. \tag{C4}$$

From Eq. (C3), the real stationary points exist only if $\varsigma_+\varsigma_- \geq 0$. After substitution of the expressions for $t_x$, $\varsigma_+$, and $\varsigma_-$ from Eqs. (C1) and (C4) into the latter inequality, we find the transmission band defined by Eq. (16).

**Appendix D.**

**Solution of the wave equation by the perturbation theory**

For sufficiently small modulation of the refractive index $\Delta n_p$ and small modulation index $|\Omega_p| \ll 1$, solution of the wave equation, Eq. (1), can be found by the perturbation theory. We rewrite Eq. (1) as

$$\left((n_0 + \Delta n(x,t))^2 E\right)_{tt} - c^2 E_{xx} = 0 \tag{D1}$$

and solve it by perturbations,

$$E(x,t) = E^{(0)}(x,t) + E^{(1)}(x,t) + E^{(2)}(x,t)\ldots, \tag{D2}$$

over $\Delta n(x,t)$. The general solution of Eq. (D1) is

$$E^{(gen)}(x,t) = E(x,t) + \Psi^+\left(t - \frac{x}{v_0}\right) + \Psi^-\left(t + \frac{x}{v_0}\right). \tag{D3}$$

where $\Psi^\pm(t)$ are arbitrary functions. The general solution of our interest, which is used in calculations of the transmission amplitude through an optical resonator corresponds to $\Psi^-(t) \equiv 0$, since it includes only optical waves propagating along a positive direction of axis $x$. In the zero, first, and second order, we have

$$n_0^2 E_{tt}^{(0)} - c^2 E_{xx}^{(0)} = 0, \tag{D4}$$

$$n_0^2 E_{tt}^{(1)} - c^2 E_{xx}^{(1)} = -2n_0 \left(\Delta n(x,t) E^{(0)}\right)_{tt}, \tag{D5}$$

$$n_0^2 E_{tt}^{(2)} - c^2 E_{xx}^{(2)} = -2n_0 \left(\Delta n^2 E^{(0)}\right)_{tt} - 2n_0 \left(\Delta n E^{(1)}\right)_{tt}. \tag{D6}$$

Here we take into account the attenuation $\alpha$ of modulation along the waveguide length setting

$$\Delta n(x,t) = \Delta n_{p0} \exp(-\alpha x) \cos\left(\omega_p\left(t - \frac{x}{v_p}\right)\right). \tag{D7}$$

We choose the zero-order solution of Eq. (D1) as

$$E^{(0)}(x,t) = \exp\left[i\omega_0\left(\frac{x}{v_0} - t\right)\right]. \tag{D8}$$

To solve Eq. (D5) with $E^{(0)}(0,t)$ defined by Eq. (D8), we separate Eq. (D5) into two equations for $E^{(1)(+)}(x,t)$ and $E^{(1)(-)}(x,t)$:

$$n_0^2 E_{tt}^{(1)(\pm)} - c^2 E_{xx}^{(1)(\pm)} = -n_0 \Delta n_{p0} \left(\exp\left(\pm i\omega_p\left(\frac{x}{v_p} - t\right) + i\omega_0\left(\frac{x}{v_0} - t\right) - \alpha x\right)\right)_{tt}. \tag{D9}$$

Then, a particular solution of Eq. (D5) vanishing for $\Delta n(x,t) \to 0$ is

$$E^{(1)}(x,t) = E^{(1)(+)}(x,t) + E^{(1)(-)}(x,t). \tag{D10}$$

Functions $E^{(1)(\pm)}(x,t)$ can be found in the form proportional to the right-hand side function of Eq. (D9):

$$E^{(1)(\pm)}(x,t) = \Delta U_0^\pm \exp\left(\pm i\omega_p\left(\frac{x}{v_p} - t\right) + i\omega_0\left(\frac{x}{v_0} - t\right) - \alpha x\right). \tag{D11}$$

Substitution Eq. (D11) into Eq. (D9) yields:

$$\Delta U_0^{\pm} = \frac{\Delta n_p v_p^2 (\omega_0 \pm \omega_p)^2}{n_0 \left[ \pm \omega_p (v_0 - v_p) + i\alpha v_0 v_p \right] \left[ 2\omega_0 v_p \pm \omega_p (v_0 + v_p) + i\alpha v_0 v_p \right]}. \tag{D12}$$

Choosing an appropriate function $\Psi^+(t)$ in Eq. (D3), we find the first order solution of Eq. (D1) satisfying the boundary condition $E^{(1)}(0,t) = 0$:

$$E^{(1)}(x,t) = E^{(0)}(x,t) \sum_{\pm} \left( \Delta U_0^{\pm} \exp(\mp i\omega_p t) W^{\pm}(x) \right), \tag{D13}$$

$$W^{\pm}(x) = \exp\left( \pm i \frac{\omega_p x}{v_p} - \alpha x \right) - \exp\left( \pm i \frac{\omega_p x}{v_0} \right). \tag{D14}$$

This solution can be directly transformed to the form presented by Eqs. (36)-(39).

Here we are only interested in the part of second order solution $E^{(2)}(x,t)$ of Eq. (D1) which contributes to the averaged over time output transmission power $|E(x,t)|^2$ at the resonance modulation frequency $\omega_p^{(res)}$ determined by Eq. (42) and under the condition $\exp(-\alpha x) \ll 1$ discussed in the main text. Direct calculations show that close to the resonance $\omega_p^{(res)}$ and under the commonly valid condition $\alpha v_0 \ll \omega_0$ the contribution of the second order solution $E^{(2)}(x,t)$ to the averaged power is much smaller than the contribution of the first order. Consequently, we can use Eq. (44) to find

$$P_{av} = \langle |E(x,t)|^2 \rangle_t \cong 1 + |U_{res}^-|^2 \tag{D15}$$

This equation is equivalent to Eq. (45) of the main text.

**Appendix E.**

**Asymptotic calculation of the optical resonator transmission spectrum**

We are interested here in finding the asymptotic expression for the amplitudes of optical frequency comb lines with large numbers $m \gg 1$ for a racetrack optical microresonator weakly coupled to a waveguide with $\kappa \ll 1$. Under the latter assumption, we have $\tau \cong 1 - \kappa^2/2$ and, thus, $\tau^n \cong \exp(-n\kappa^2/2)$. For small propagation losses, $\omega_0 T \eta_0 / n_0 \ll 1$, at exact optical resonance, $\omega_0 T = 2\pi N_0$ with integer $N_0$, and close to the modulation resonance, $\omega_p T = 2\pi N_p + \Delta \omega_p T$ with $\Delta \omega_p T \ll 1$ and integer $N_p$, the expression for the absolute values of the comb spectral amplitudes $U_m^{(c)}$ in Eq. (51) can be written as

$$\left| U_m^{(c)} \right| = \kappa^2 \exp\left[ -m \frac{\omega_p}{2\omega_0} \left( \frac{v_0}{v_p} - 3 \right) \right] \times \left| \sum_{n=0}^{\infty} J_m(\bar{\sigma}_n \Omega_p) \exp\left[ n \left( -\frac{im}{2} \Delta \omega_p T - \rho \right) \right] \right|,$$

$$\bar{\sigma}_n = \frac{2}{\Delta \omega_p T} \sin\left( \frac{n}{2} \Delta \omega_p T \right), \quad \rho = \frac{\eta_0}{n_0} \omega_0 T + \frac{\kappa^2}{2}, \quad T = \frac{2L}{v_0}. \tag{E1}$$

To calculate the sum in Eq. (E1), we use the asymptotic expression for the Bessel function:

$$J_m(x)\big|_{|m|,|x|,|x|-|m| \gg 1} \cong \left( \frac{2}{\pi} \right)^{1/2} (x^2 - m^2)^{-1/4} \cos\left( (x^2 - m^2)^{1/2} - m \arccos\left( \frac{m}{x} \right) - \frac{\pi}{4} \right). \tag{E2}$$

For $m\Delta\omega_p T/2 \ll 1$, we can replace the sum in Eq. (E1) by an integral. Then, presenting the cosine in Eq. (E2) as a sum of exponents we find

$$\left|U_m^{(c)}\right| = \frac{\kappa^2}{(2\pi)^{1/2}} \exp\left[-m\frac{\omega_p}{\omega_0}\left(\frac{v_0}{v_p} - 3\right)\right] \left|\sum_\pm \left[\int_0^\infty dn \left(\bar{\sigma}_n^2 \Omega_p^2 - m^2\right)^{-1/4} \exp\left(i\varphi_{m,n}^\pm - n\rho\right)\right]\right|,$$

(E3)

$$\varphi_{m,n}^\pm = \pm\left(\bar{\sigma}_n^2 \Omega_p^2 - m^2\right)^{1/2} \mp m\arccos\left(\frac{m}{\bar{\sigma}_n \Omega_p}\right) - \frac{mn}{2}\Delta\omega_p T \mp \frac{\pi}{4}.$$

We calculate the integral over $n$ in Eq. (E3) by the stationary phase method. Assuming that for the stationary points $n = n_m^\pm$ corresponding to signs + and − in Eq. (E3) we have $n_m^\pm \alpha \sim 1$ (see below), we find these points from the equation:

$$\frac{\partial \varphi_{m,n}^\pm}{\partial n} = \pm \frac{\Delta\omega_p T}{2}\left(\frac{4\Omega_p^2}{\Delta\omega_p^2 T^2}\sin^2\left(\frac{n_m^\pm}{2}\Delta\omega_p T\right) - m^2\right)^{1/2} \cot\left(\frac{n_m^\pm}{2}\Delta\omega_p T\right) - \frac{m}{2}\Delta\omega_p T = 0.$$

(E4)

This equation can be simplified to

$$\sin\left(n_m^\pm \Delta\omega_p T\right) = \pm \Upsilon_m,$$

(E5)

where

$$\Upsilon_m = \frac{m\Delta\omega_p T}{\Omega_p}.$$

(E6)

The condition of existence of stationary points $|\Upsilon_m| \leq 1$ following from Eqs. (E5) and (E6) determines the transmission band

$$-\frac{\Delta\omega_B}{2} + \omega_0 \leq \omega \leq \frac{\Delta\omega_B}{2} + \omega_0$$

(E7)

with the bandwidth

$$\Delta\omega_B = \frac{2|\Omega_p|\omega_p}{|\Delta\omega_p|T}.$$

(E8)

Next, we find

$$\frac{\partial^2 \varphi_{m,n}^\pm}{\partial n^2} = \pm\frac{\Delta\omega_p T \Omega_p}{2} \frac{1 - 2\left(\Theta_m^\pm\right)^2}{\left(\left(\Theta_m^\pm\right)^2 - \frac{\mu_m^2}{4}\right)^{1/2}}, \quad \Theta_m^\pm = \sin\left(\frac{n_m^\pm}{2}\Delta\omega_p T\right).$$

(E9)

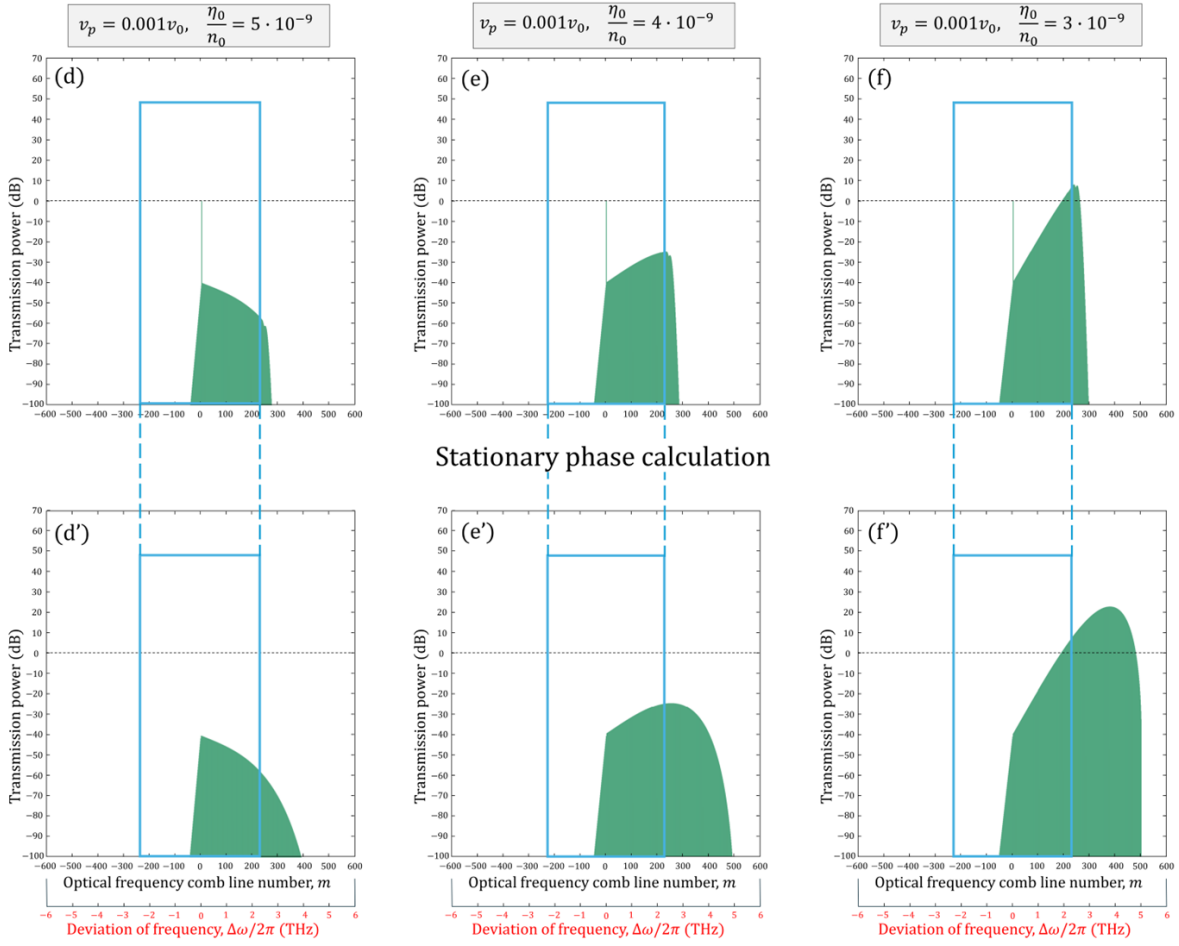

**Fig. 13.** Transmission spectra calculated exactly using Eq. (E1) and presented in Fig. 9 of the main text and those calculated using Eq. (E15) derived by the stationary phase method. The frequency range where the stationary phase method is valid and results in the spectrum accurately coinciding with the exact calculations based on Eq. (51) is outlined by rectangles.

Applying the stationary phase method to Eq. (E1) we find

$$\left|U_m^{(c)}\right| = \frac{\kappa^2}{\Omega_p} \exp\left[-m\frac{\omega_p}{2\omega_0}\left(\frac{v_0}{v_p}-3\right)\right] \left|\sum_{\pm,n_m^\pm}\left[\pm\left(1-2\left(\Theta_m^\pm\right)^2\right)^{-1/2} \exp\left(i\varphi_{m,n_m^\pm}^\pm - n_m^\pm \rho\right)\right]\right|. \quad \text{(E10)}$$

For determinacy, we assume that $\Delta\omega_p T > 0$ and $\Omega_p > 0$. Since the integration in Eq. (E3) is over the positive $n$ only, we are looking for the positive solutions of Eq. (E5), which are defined as:

$$n_{m,\pm,N_+}^+ = \frac{1}{\Delta\omega_p T}\left(\pm\arcsin(\Upsilon_m) + \frac{\pi}{2}(1\mp 1) + 2\pi N_+\right), \quad N_+ = 0,1,\dots, \quad \text{(E11)}$$

$$n_{m,\pm,N_+}^- = \frac{1}{\Delta\omega_p T}\left(\mp\arcsin(\Upsilon_m) + \frac{\pi}{2}(1\pm 1) + 2\pi(N_- + 1)\right), \quad N_- = 0,1,\dots. \quad \text{(E12)}$$

Crucially, due to the attenuation factor $\exp(-n_m^{\pm}\rho)$, the absolute values of terms in the sum of Eq. (E10) can be significantly different. Two smallest positive values of stationary points are

$$n_{m,\pm,0}^{+} = \frac{1}{\Delta\omega_p T}\left(\pm\arcsin(|\Upsilon_m|) + \frac{\pi}{2}(1\mp 1)\right), \tag{E13}$$

while the contribution of other stationary points to this sum is at least $\exp(\pi\rho/\Delta\omega_p T)$ times smaller than the largest contribution of these two points. Thus, we can ignore the contribution of the other stationary points if

$$\exp\left(\frac{\pi\rho}{\Delta\omega_p T}\right) \gg 1. \tag{E14}$$

Under the condition of Eq. (E14), the expression for the comb spectral amplitudes $U_m^{(c)}$ in Eq. (51) is reduced to

$$\left|U_m^{(c)}\right| = \frac{\kappa^2}{\Omega_p}\left(1-\Upsilon_m^2\right)^{-1/4}\exp\left[-m\frac{\omega_p}{\omega_0}\left(\frac{v_0}{v_p}-3\right)\right]\times\left|\exp\left(-n_{m,+}^{+}\alpha + i\varphi_{m,n_{m,+,0}^{+}}^{+}\right) - \exp\left(-n_{m,-}^{+}\alpha + i\varphi_{m,n_{m,-,0}^{+}}^{+}\right)\right|. \tag{E15}$$

Fig. 13 demonstrates a reasonable agreement between the output spectral power $\left|U_m^{(c)}\right|^2$ calculated using exact Eq. (51) (top plots) and using asymptotic Eq. (E15) (bottom plots). Here, the top plots replicate Figs. 9(d), (e), and (f) of the main text.